\documentclass{aa}
\usepackage{graphicx}
\usepackage{bm}

\newcommand{\EQ}{\begin{equation}}
\newcommand{\EE}{\end{equation}}
\newcommand{\EQA}{\begin{eqnarray}}
\newcommand{\EEA}{\end{eqnarray}}

\newcommand{\pd}{\partial}

\newcommand{\DIV}{\vec{\nabla} \cdot }

\newcommand{\ve}[1]{\boldsymbol{#1}}
\newcommand{\mean}[1]{\overline{#1}}
\newcommand{\meanv}[1]{\overline{\bm #1}}
\newcommand{\cst}{c_{\rm s}^2}
\newcommand{\nut}{\nu_{\rm t}}
\newcommand{\urms}{u_{\rm rms}}

\newcommand{\kef}{k_{\rm f}}
\newcommand{\tauc}{\tau_{\rm c}}
\newcommand{\tauto}{\tau_{\rm to}}
\newcommand{\St}{{\rm St}}
\newcommand{\Co}{{\rm Co}}
\newcommand{\Cost}{\Omega_\star}

\newcommand{\qxx}{Q_{xx}}
\newcommand{\qyy}{Q_{yy}}
\newcommand{\qzz}{Q_{zz}}
\newcommand{\qxy}{Q_{xy}}
\newcommand{\qxz}{Q_{xz}}
\newcommand{\qyz}{Q_{yz}}
\newcommand{\qij}{Q_{ij}}
\newcommand{\Omx}{\Omega_x}
\newcommand{\Omz}{\Omega_z}
\def\onethird{{\textstyle{1\over3}}}
\def\onehalf{{\textstyle{1\over2}}}
\def\threefourths{{\textstyle{3\over4}}}

\begin{document}

\authorrunning{K\"apyl\"a \& Brandenburg}
\titlerunning{Lambda effect from forced turbulence simulations}

   \title{Lambda effect from forced turbulence simulations}

   \author{P. J. K\"apyl\"a
	  \inst{1,2}
          \and
          A. Brandenburg
	  \inst{2}
	  }

   \offprints{P. J. K\"apyl\"a\\
	  \email{petri.kapyla@helsinki.fi}
	  }

   \institute{Observatory, PO Box 14, FI-00014 University of Helsinki, 
              Finland
         \and NORDITA, AlbaNova University Center, Roslagstullsbacken
              23, SE-10691 Stockholm, Sweden}

   \date{Received 19 November 2007 / Accepted 26 May 2008}

   \abstract{}%
   {We determine the components of the $\Lambda$-effect tensor that quantifies
    the contributions to the turbulent momentum transport
    even for uniform rotation.}
   {Three-dimensional numerical simulations are used to study
     turbulent transport in triply periodic cubes under the influence
     of rotation and anisotropic forcing. Comparison is made with
     analytical results obtained via the so-called minimal
     tau-approximation.}%
   {In the case where the turbulence intensity in the vertical
     direction dominates, the vertical stress is always negative.
     This situation is expected to occur in stellar convection zones.
     The horizontal component of the stress
     is weaker and exhibits a
     maximum at latitude $30\degr$ --- regardless of how rapid
     the rotation is. The minimal tau-approximation captures many of the
     qualitative features of the numerical results, provided the relaxation
     time tau is close to the turnover time, i.e.\ the Strouhal
     number is of order unity.
  }{}


   \maketitle

%________________________________________________________________

\section{Introduction}
Differential rotation plays a crucial role in dynamo processes
that sustain large-scale magnetic activity in stars like the
Sun (e.g.\ Moffatt \cite{Moffatt1978}; Krause \& R\"adler
\cite{KrauRaed1980}). The internal rotation of the Sun is familiar
from helioseismology (e.g.\ Thompson et al.\ \cite{Thompsonea2003}), but
the processes sustaining the observed rotation profile
are not understood well.
The angular momentum balance of a rotating star is determined by the
conservation equation%
\EQ%
\frac{\pd}{\pd t} (\rho s^2 \mean\Omega) + \bm\nabla \cdot (\rho s^2 \mean\Omega\;
\meanv{U} + \rho s \mean{u_\phi \bm{u}}) = 0\;,%
\EE%
where $\meanv{U}$ is the meridional flow, $s$
the cylindrical radius,
$\rho$ the density (neglecting however its fluctuations),
$\mean\Omega=\mean{U}_\phi/s$ the local angular velocity, and
$\mean{u_\phi \bm{u}}$ the zonal component of the Reynolds
stress. Overbars denote averages over the azimuthal
direction.

The meridional flow can also be directly affected by the Reynolds
stresses (e.g.\ R\"udiger \cite{Ruediger1989}), but it is more strongly
determined by the baroclinic term that arises if the isocontours of
density and pressure do not coincide. Such a configuration can appear
because of latitude-dependent turbulent heat fluxes that arise naturally
in rotating convection (e.g.\ R\"udiger \cite{Ruediger1982}; Pulkkinen
et al.\ \cite{Pulkkinenea1993}; K\"apyl\"a et
al.\ \cite{Kaepylaeea2004}; R\"udiger et al.\ \cite{Ruedigerea2005a}) or
from a subadiabatic tachocline (Rempel \cite {Rempel2005}) which is
likely to occur in the Sun (Rempel \cite{Rempel2004}; K\"apyl\"a et
al.\ \cite{Kaepylaeea2006c}). The flows due to thermodynamic
effects are likely to be needed to avoid the Taylor--Proudman
balance in the Sun (e.g.\ Durney \cite{Durney1989};
Brandenburg et al.\ \cite{Brandenburgea1992};
Kitchatinov \& R\"udiger \cite{KitRued1995}; Rempel \cite{Rempel2005}). 
The overall importance
of the meridional flow in the angular momentum balance of the Sun is,
however, still unclear since no definite observational information
about it is available below roughly 20\,Mm depth (e.g.\ Zhao \&
Kosovichev \cite{ZhaoKoso2004}).

Although not much more is known about the Reynolds stresses from
observations, already this limited knowledge can be used to gain
insight into the theory of turbulent transport. Solar surface
observations indicate that there is an equatorward flux of angular
momentum, as described by the Reynolds stress component $Q_{\theta \phi}
= \overline{u_\theta u_\phi}$, of several
$10^3$\,m$^2$\,s$^{-2}$ in the latitude range where sunspots are
observable (e.g.\ Ward \cite{Ward1965}; Pulkkinen \& Tuominen
\cite{PulkTuo1998}; Stix \cite{Stix2002}). In mean-field theory the
simplest approximation that can be made concerning the Reynolds
stresses is to assume them proportional to the gradient of mean velocity
(the Boussinesq ansatz):
\begin{equation}
Q_{ij} \equiv \overline{u_i u_j} = - \mathcal{N}_{ijkl} \overline{U}_{k,l}\;.
\end{equation}
In the Sun this ansatz turns out to be insufficient because the
observed $Q_{\theta \phi}$ and solar surface differential rotation
profile indicate that the turbulent viscosity is negative. Thus, in
analogy to mean-field dynamo theory, additional contributions to the
Reynolds stress were conjectured to appear (e.g.\ Wasiuty$\acute{\rm n}$ski
\cite{Wasiutynski1946}; Krause \& R\"udiger \cite{KrauseRued1974}),
leading to the present description
\begin{equation}
Q_{ij} =  \Lambda_{ijk} \mean\Omega_k - \mathcal{N}_{ijkl} \overline{U}_{k,l}\;,
\end{equation}
where $\Lambda_{ijk}$ is a third-rank tensor
describing the $\Lambda$-effect that contributes to
the Reynolds stress even in the case of rigid rotation.
These terms are often referred to as ``non-diffusive'' contributions
to the Reynolds stress.
The zonal components of the stress
can be written in the form (e.g.\ Stix \cite{Stix2002})%
\EQA%
Q_{\theta \phi} &=& \Lambda_{\rm H} \cos \theta\, \mean\Omega - \nut \sin \theta \frac{\pd \mean\Omega}{\pd \theta}\;,\\%
Q_{r \phi} &=& \Lambda_{\rm V} \sin \theta\, \mean\Omega - \nut (1 -
\epsilon)  r \sin
\theta \frac{\pd \mean\Omega}{\pd r}\;,%
\EEA%
where $\Lambda_{\rm H}$ and $\Lambda_{\rm V}$ describe the
non-diffusive transport and where $\nut$ is the turbulent
viscosity. The factor $1-\epsilon$ in the latter equation indicates
that the turbulent viscosity can be anisotropic. Furthermore, the
coefficients $\Lambda_{\rm H}$, $\Lambda_{\rm V}$, and $\nut$ can vary as functions of the spatial coordinates.

Much effort has been devoted to determining Reynolds stresses
from convection simulations (Hathaway \& Somerville
\cite{HathaSomer1983}; Pulkkinen et al.\ \cite{Pulkkinenea1993};
Rieutord et al.\ \cite{Rieutordea1994}; Brummell et
al.\ \cite{Brummellea1998}; Chan \cite{Chan2001}; K\"apyl\"a et
al.\ \cite{Kaepylaeea2004}; R\"udiger et al.\ \cite{Ruedigerea2005b};
Hupfer et al.\ \cite{Hupferea2005}, \cite{Hupferea2006}; Giesecke
\cite{Giesecke2007}). These studies have confirmed the existence of
the $\Lambda$-effect and revealed some surprising results that are at
odds with theoretical considerations (e.g.\ Kitchatinov \& R\"udiger
\cite{KitRued1993}, \cite{KitRued2005}) derived under the second-order
correlation approximation (SOCA). The discrepancies are most prominent
in the rapid rotation regime, $\Cost \approx 10$, where
\EQ%
\Cost = 2\, \Omega_0 \tauto\;, \label{equ:Corioliscom}%
\EE%
is the Coriolis (or the inverse Rossby) number. Here, $\Omega_0$ is
the angular momentum-averaged
rotation rate and $\tauto$ the convective turnover time.
In the solar convection zone, $\Cost$ varies between 10$^{-3}$ near the
surface to ten or more in the deep layers. Convection simulations in
the latter regime show that the horizontal angular momentum flux is
directed toward the equator, corresponding to positive $Q_{\theta
  \phi}$ in the northern hemisphere, and that it peaks very sharply near 
the equator (Chan
\cite{Chan2001}; K\"apyl\"a et al.\ \cite{Kaepylaeea2004}; Hupfer et
al.\ \cite{Hupferea2005}). On the other hand, the vertical stress $Q_{r
  \phi}$ can be directed outward (K\"apyl\"a et
al.\ \cite{Kaepylaeea2004}), contradicting the theory for
vertically dominated turbulence (e.g.\ Biermann \cite{Biermann1951}; 
R\"udiger \cite{Ruediger1980},
\cite{Ruediger1989}). So far, these results remain without proper
explanation.

Often the Reynolds stress realized in the simulation is taken to solely 
represent
the $\Lambda$-effect. This approach seems like a reasonable starting
point but in an inhomogeneous system large scale mean flows are
generated when rotation becomes important. These flows affect the Reynolds 
stresses via the turbulent
viscosity. Furthermore, in the presence of stratification, heat fluxes
can also significantly affect the stresses (Kleeorin \& Rogachevskii
\cite{KleeRoga2006}). In the present study we simplify the situation
as much as possible in order to disentangle the effect of the turbulent
velocity field from other effects. Thus we neglect stratification and
heat fluxes by adopting a periodic isothermal setup.
Turbulence is driven by external forcing, which
provides clear scale separation between the turbulent eddies and the
system size, which is typically not achieved in convection
simulations.
Further insight
is sought from comparison of simple analytical closure models with
numerical data.

Preliminary results on the $\Lambda$-effect are
presented in K\"apyl\"a \& Brandenburg (\cite{KaBr2007}). In the
present paper, numerical datasets covering a significantly larger part
of the parameter space are analyzed, and a more thorough study of the
validity and results of the minimal tau-approximation are presented.

The remainder of the paper is organized as follows.
Section~\ref{sec:model} summarizes the model and the methods of the
study, and in Sects.~\ref{sec:results} and \ref{sec:conclusions} the
results and the conclusions are given.
\vfill

\section{The model and methods}
\label{sec:model}
\subsection{Basic equations}
We model compressible hydrodynamic turbulence in a triply periodic
cube of size $(2\pi)^3$. The gas obeys an isothermal equation of state
characterized by a constant speed of sound, $c_{\rm s}$. The
continuity and Navier-Stokes equations read
\begin{equation}
\frac{\rm{D} \ln \rho}{\rm{D} t} = - \vec{\nabla} \cdot \vec{U}\;,
\end{equation}
\begin{equation}
\frac{\rm{D} \vec{U}}{\rm{D} t} = - c_{\rm s}^{2} \vec{\nabla} \ln \rho - 2\, \bm{\Omega} \times \bm{U} + \vec{f}_{\rm visc} + \vec{f}_{\rm force}\;,\label{equ:NS}
\end{equation}
where $\rm{D}/\rm{D} t=\partial/\partial t+\vec{U}\cdot\bm{\nabla}$
denotes the advective derivative, $\vec{U}$ is the velocity, $\rho$
the density, $\vec{f}_{\rm visc}$ the viscous force, and
$\vec{f}_{\rm force}$ the forcing function. Due to the periodic
boundaries, mass is conserved and the average density retains its
initial value $\overline{\rho} = \rho_0$ at all times. Compressibility
is retained but we consider low Mach number flows, $\urms/c_{\rm s}
\approx 0.1-0.2$. The viscous force is given by
\begin{equation}
\bm{f}_{\rm visc} = \nu \Big(\nabla^2 \bm{U} +\onethird \vec{\nabla} \DIV \bm{U} + 2\, \bm{\mathsf{S}} \cdot \vec{\nabla} \ln \rho \Big)\;,
\end{equation}
where $\nu$ is the kinematic viscosity and
\begin{equation}
\mathsf{S}_{ij} = \onehalf \bigg(\frac{\pd U_i}{\pd x_j} + \frac{\pd U_j}{\pd x_i} \bigg) - \onethird \delta_{ij} \frac{\pd U_k}{\pd x_k}\;,
\end{equation}
the traceless rate of strain tensor.
The forcing function $\bm{f}_{\rm force}$ is given by
\begin{eqnarray}
\bm{f}(\bm{x},t) = {\rm Re} \{\bm{N} \cdot \bm{f}_{\bm{k}(t)} \exp [i \bm{k}(t)
  \cdot \bm{x} - i \phi(t) ] \}\;,
\end{eqnarray}
where $\bm{x}$ is the position vector, ${\bm N} = \bm{f}^{(0)} c_{\rm
  s} (k c_{\rm s}/\delta t)^{1/2}$ is a normalization factor, and
$\bm{f}^{(0)}$ describes the direction-dependent amplitude of the
forcing (see below), $k = |\bm{k}|$, $\delta t$ is the length of
the time step, and $-\pi < \phi(t) < \pi$ a random delta-correlated
phase. The vector $\bm{f}_{\bm{k}}$ is given by
\begin{eqnarray}
\bm{f}_{\ve{k}} = \frac{\bm{k} \times \hat{\bm{e}}}{\sqrt{\bm{k}^2 - (\bm{k}
    \cdot \hat{\bm{e}})^2}}\;,
\end{eqnarray}
where $\hat{\bm{e}}$ is an arbitrary unit vector. Thus, $\bm{f}_{\bm{k}}$
describes nonhelical transversal waves with $|\bm{f}_{\ve{k}}|^2 = 1$,
where ${\bm k}$ is chosen randomly from a predefined range in the vicinity
of the average wavenumber $\kef/k_1 = 5$ at each time step, where $k_1$ 
is the wavenumber corresponding to the domain size,
and $\kef$ the wavenumber of the energy-carrying scale. 

To make the resulting turbulence anisotropic the forcing
amplitude depends on direction
\begin{eqnarray}
f_i^{\rm (aniso)} = (\delta_{ij} f_0 + \hat{z}_i \hat{z}_j \cos^2 \Theta_{\bm k} f_1) f_j^{\rm (iso)}  \label{equ:forceaniso}
\end{eqnarray}
where $f_0$ and $f_1$ are the amplitudes of the isotropic and
anisotropic parts of the forcing, $\hat{\bm z}$ the unit vector in the
vertical direction, and $\Theta_{\bm k}$ the angle between the
vertical direction and the wave vector ${\bm k}$.
In the following we assume $f_1 \gg f_0$, but even then
the amount of anisotropy depends on the Reynolds
number and is in any case only quite modest (see,
e.g.\ Table~\ref{tab:Re12}).

At this point it is important to emphasize that our forcing function is
designed to capture the effects that lead to a finite $\Lambda$-effect.
The implementation of anisotropy should therefore be a simple one.
In stars, anisotropy is produced by stratification and convection.
Our goal is clearly not to simulate properties of convection other
than its tendency to make the turbulence anisotropic.

The numerical computations were made with the {\sc
  Pencil-Code}\footnote{\texttt{http://www.nordita.org/software/pencil-code/}},
which uses sixth-order accurate finite differences in space, and a
third-order accurate time-stepping scheme (Brandenburg \& Dobler
\cite{BranDobler2002}; Brandenburg
\cite{Brandenburg2003}). Resolutions up to $256^3$ grid points were used in the
simulations.

\subsection{Nondimensional units}

In the following we use non-dimensional variables by setting
\begin{eqnarray}
c_{\rm s}=k_1=\rho_0=1.
\end{eqnarray}
This means that the units of length, time, and density are
\begin{eqnarray}
[x]=k_1^{-1},\;\;
[t]=(c_{\rm s}k_1)^{-1},\;\;
[\rho]=\rho_0.
\end{eqnarray}
However, in most of the plots we present the results in explicitly
non-dimensional form using the quantities above.

\subsection{Coordinate system, averaging, and error estimates}
The simulated domain is thought to represent a small rectangular
portion of a spherical body of gas. We choose $(x,y,z)$ to correspond
to $(\theta,\phi,r)$ of spherical coordinates. With this choice the
rotation vector can be written as
\begin{equation}
\bm{\Omega} = \Omega_0(-\sin \theta,0,\cos \theta)^T,
\label{OmegaDef}
\end{equation}
where $\theta$ is the angle between the rotation axis and the
local vertical direction, i.e.\ the colatitude.

\begin{figure}[t]
\centering
\includegraphics[width=0.5\textwidth]{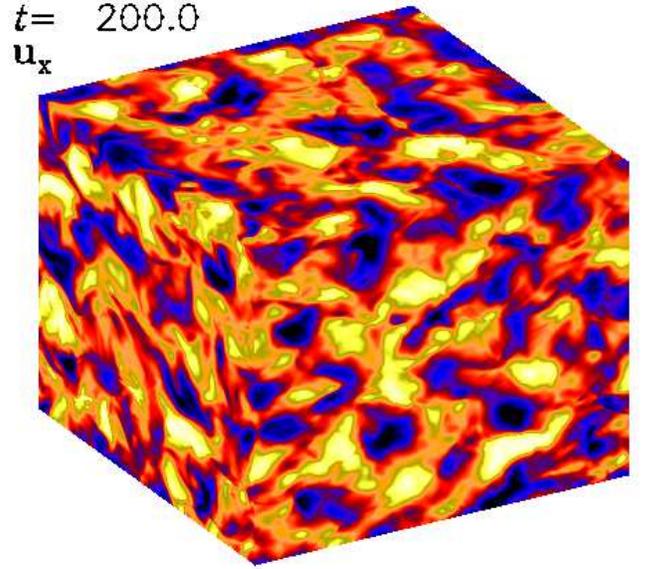}
\caption{$U_x$ at the periphery of the simulation domain from a slowly
  rotating run with $\Co \approx 0.3$, $\kef/k_1 = 5$, and
  $\theta=0\degr$, resolution $256^3$.}
\label{fig:img_0099}
\end{figure}

Since the turbulence is homogeneous, volume averages are employed and denoted by overbars. 
An additional time average over the statistically saturated state of
the calculation is also taken.
We define the Coriolis number as%
\EQA%
\Co = \frac{2\, \Omega_0}{\urms \kef}\;, \label{equ:Coriolis}%
\label{equ:Corioliscomnew}%
\EEA%
where $\urms$ is the rms-velocity.
In comparison to the commonly used
definition (\ref{equ:Corioliscom}), definition (\ref{equ:Corioliscomnew})
is smaller by a factor of $2\pi$, i.e. $\Co = \Cost/2\pi$. 
The other dimensionless number relevant in this
study is the Reynolds number based on the forcing scale%
\EQA%
{\rm Re} &=& \frac{\urms}{\nu \kef}\;.%\\
\EEA%
See Fig.~\ref{fig:img_0099} for a snapshot from a typical run with
${\rm Re} \approx 60$.

Errors are estimated by dividing the time series into three equally
long parts and computing mean values for each part individually. The
largest departure from the mean value computed for the whole time
series is taken to represent the error.

\onltab{1}{
   \begin{table*}
   \centering
   \caption[]{Summary of the turbulence anisotropies and normalized Reynolds stresses, $\tilde{Q}_{ij} = Q_{ij}/\urms^2$, for the ${\rm Re} \approx 12 \ldots 14$ calculations. $\kef/k_1=5$, $\nu = 2 \cdot 10^{-3}$, and grid resolution $64^3$ was used in all runs.}
      \vspace{-0.5cm}
      \label{tab:Re12}
     $$
         \begin{array}{p{0.055\linewidth}cccccccrcrrr}
           \hline
           \noalign{\smallskip}
           Run      & \Omega_0  & \theta & \Co & \urms & \tilde{Q}_{xx} & \tilde{Q}_{yy} & \tilde{Q}_{zz} & A_{\rm H} & A_{\rm V} & \tilde{Q}_{xy} & \tilde{Q}_{xz} & \tilde{Q}_{yz} \\
           \noalign{\smallskip}
           \hline
           \noalign{\smallskip}
           64a1  & 0.02 &  0\degr &  0.06 & 0.127 & 0.236 & 0.240 & 0.525 & 0.002 & -0.571 &-0.001 &-0.000 & 0.003 \\
           64a2  & 0.02 & 15\degr &  0.06 & 0.127 & 0.239 & 0.242 & 0.523 & 0.003 & -0.567 &-0.001 &-0.002 &-0.004 \\
           64a3  & 0.02 & 30\degr &  0.06 & 0.127 & 0.237 & 0.243 & 0.523 & 0.004 & -0.564 &-0.001 &-0.003 &-0.001 \\
           64a4  & 0.02 & 45\degr &  0.06 & 0.127 & 0.238 & 0.244 & 0.522 & 0.005 & -0.561 &-0.000 &-0.004 &-0.014 \\
           64a5  & 0.02 & 60\degr &  0.06 & 0.127 & 0.237 & 0.246 & 0.521 & 0.008 & -0.559 &-0.000 &-0.003 &-0.018 \\
           64a6  & 0.02 & 75\degr &  0.06 & 0.127 & 0.237 & 0.246 & 0.521 & 0.010 & -0.559 &-0.001 &-0.002 &-0.021 \\
           64a7  & 0.02 & 90\degr &  0.06 & 0.127 & 0.236 & 0.247 & 0.521 & 0.010 & -0.558 &-0.001 &-0.001 &-0.021 \\
           \hline
           \noalign{\smallskip}
           64a8  & 0.05 &  0\degr &  0.15 & 0.127 & 0.244 & 0.246 & 0.514 & 0.003 & -0.537 & 0.000 & 0.001 & 0.002 \\
           64a9  & 0.05 & 15\degr &  0.15 & 0.127 & 0.244 & 0.248 & 0.512 & 0.003 & -0.532 &-0.000 &-0.005 &-0.007 \\
           64a10 & 0.05 & 30\degr &  0.15 & 0.127 & 0.243 & 0.250 & 0.511 & 0.006 & -0.529 & 0.002 &-0.010 &-0.018 \\
           64a11 & 0.05 & 45\degr &  0.15 & 0.127 & 0.242 & 0.253 & 0.509 & 0.011 & -0.522 & 0.004 &-0.012 &-0.027 \\
           64a12 & 0.05 & 60\degr &  0.15 & 0.127 & 0.239 & 0.258 & 0.507 & 0.019 & -0.516 & 0.005 &-0.009 &-0.035 \\
           64a13 & 0.05 & 75\degr &  0.15 & 0.127 & 0.236 & 0.263 & 0.505 & 0.027 & -0.512 & 0.004 &-0.005 &-0.040 \\
           64a14 & 0.05 & 90\degr &  0.15 & 0.127 & 0.234 & 0.265 & 0.505 & 0.032 & -0.510 &-0.000 & 0.000 &-0.042 \\
           \hline
           \noalign{\smallskip}
           64a15 &  0.1 &  0\degr &  0.31 & 0.126 & 0.255 & 0.258 & 0.490 & 0.003 & -0.468 &-0.000 & 0.002 & 0.002 \\
           64a16 &  0.1 & 15\degr &  0.31 & 0.127 & 0.256 & 0.257 & 0.490 & 0.001 & -0.468 & 0.001 &-0.009 &-0.004 \\
           64a17 &  0.1 & 30\degr &  0.31 & 0.127 & 0.257 & 0.257 & 0.490 &-0.000 & -0.466 & 0.005 &-0.019 &-0.014 \\
           64a18 &  0.1 & 45\degr &  0.31 & 0.127 & 0.254 & 0.261 & 0.489 & 0.007 & -0.462 & 0.011 &-0.022 &-0.028 \\
           64a19 &  0.1 & 60\degr &  0.31 & 0.127 & 0.247 & 0.272 & 0.484 & 0.025 & -0.448 & 0.014 &-0.017 &-0.039 \\
           64a20 &  0.1 & 75\degr &  0.31 & 0.127 & 0.239 & 0.284 & 0.480 & 0.045 & -0.437 & 0.010 &-0.008 &-0.045 \\
           64a21 &  0.1 & 90\degr &  0.31 & 0.127 & 0.235 & 0.289 & 0.479 & 0.054 & -0.434 &-0.000 & 0.000 &-0.047 \\
           \hline
           \noalign{\smallskip}
           64a22 &  0.2 &  0\degr &  0.62 & 0.127 & 0.278 & 0.280 & 0.446 & 0.002 & -0.333 &-0.001 & 0.001 &-0.000 \\
           64a23 &  0.2 & 15\degr &  0.62 & 0.127 & 0.277 & 0.276 & 0.451 &-0.001 & -0.350 & 0.001 &-0.008 & 0.005 \\
           64a24 &  0.2 & 30\degr &  0.62 & 0.127 & 0.276 & 0.268 & 0.460 &-0.008 & -0.376 & 0.005 &-0.021 &-0.002 \\
           64a25 &  0.2 & 45\degr &  0.62 & 0.127 & 0.273 & 0.268 & 0.462 &-0.005 & -0.382 & 0.013 &-0.026 &-0.015 \\
           64a26 &  0.2 & 60\degr &  0.61 & 0.128 & 0.267 & 0.280 & 0.457 & 0.012 & -0.367 & 0.019 &-0.021 &-0.027 \\
           64a27 &  0.2 & 75\degr &  0.61 & 0.128 & 0.255 & 0.298 & 0.451 & 0.043 & -0.350 & 0.013 &-0.010 &-0.035 \\
           64a28 &  0.2 & 90\degr &  0.61 & 0.128 & 0.251 & 0.303 & 0.450 & 0.053 & -0.345 &-0.001 &-0.000 &-0.036 \\
           \hline
           \noalign{\smallskip}
           64a29 &  0.5 &  0\degr &  1.47 & 0.133 & 0.303 & 0.301 & 0.400 &-0.003 & -0.197 &-0.001 & 0.003 & 0.000 \\
           64a30 &  0.5 & 15\degr &  1.48 & 0.132 & 0.299 & 0.295 & 0.410 &-0.004 & -0.225 & 0.000 &-0.009 & 0.005 \\
           64a31 &  0.5 & 30\degr &  1.49 & 0.131 & 0.296 & 0.286 & 0.422 &-0.010 & -0.262 & 0.003 &-0.019 & 0.005 \\
           64a32 &  0.5 & 45\degr &  1.50 & 0.131 & 0.290 & 0.283 & 0.431 &-0.007 & -0.288 & 0.008 &-0.027 &-0.004 \\
           64a33 &  0.5 & 60\degr &  1.51 & 0.130 & 0.286 & 0.289 & 0.430 & 0.003 & -0.285 & 0.012 &-0.022 &-0.013 \\
           64a34 &  0.5 & 75\degr &  1.50 & 0.130 & 0.279 & 0.303 & 0.422 & 0.024 & -0.262 & 0.007 &-0.011 &-0.018 \\
           64a35 &  0.5 & 90\degr &  1.50 & 0.131 & 0.274 & 0.305 & 0.422 & 0.035 & -0.261 &-0.001 &-0.000 &-0.018 \\
           \hline
           \noalign{\smallskip}
           64a36 &  1.0 &  0\degr &  2.78 & 0.141 & 0.310 & 0.310 & 0.384 &-0.000 & -0.148 & 0.001 & 0.003 & 0.001 \\
           64a37 &  1.0 & 15\degr &  2.80 & 0.140 & 0.306 & 0.308 & 0.390 & 0.002 & -0.166 & 0.000 &-0.008 & 0.003 \\
           64a38 &  1.0 & 30\degr &  2.84 & 0.138 & 0.303 & 0.296 & 0.405 &-0.007 & -0.212 & 0.001 &-0.004 & 0.002 \\
           64a39 &  1.0 & 45\degr &  2.88 & 0.136 & 0.292 & 0.299 & 0.412 & 0.007 & -0.233 & 0.004 &-0.019 &-0.003 \\
           64a40 &  1.0 & 60\degr &  2.90 & 0.135 & 0.288 & 0.302 & 0.414 & 0.014 & -0.238 & 0.007 &-0.012 &-0.006 \\
           64a41 &  1.0 & 75\degr &  2.89 & 0.136 & 0.274 & 0.316 & 0.414 & 0.042 & -0.239 & 0.006 &-0.004 &-0.009 \\
           64a42 &  1.0 & 90\degr &  2.87 & 0.137 & 0.256 & 0.320 & 0.428 & 0.065 & -0.279 & 0.000 & 0.002 &-0.009 \\
           \hline
           \noalign{\smallskip}
           64a43 &  2.0 &  0\degr &  5.26 & 0.149 & 0.293 & 0.294 & 0.417 & 0.001 & -0.247 &-0.001 & 0.001 & 0.001 \\
           64a44 &  2.0 & 15\degr &  5.27 & 0.149 & 0.298 & 0.300 & 0.406 & 0.002 & -0.214 & 0.001 &-0.028 &-0.001 \\
           64a45 &  2.0 & 30\degr &  5.41 & 0.145 & 0.311 & 0.310 & 0.384 &-0.001 & -0.147 & 0.002 &-0.030 &-0.003 \\
           64a46 &  2.0 & 45\degr &  5.41 & 0.144 & 0.293 & 0.318 & 0.393 & 0.024 & -0.174 & 0.003 &-0.014 &-0.004 \\
           64a47 &  2.0 & 60\degr &  5.46 & 0.144 & 0.264 & 0.337 & 0.404 & 0.073 & -0.206 & 0.005 & 0.010 &-0.006 \\
           64a48 &  2.0 & 75\degr &  5.46 & 0.144 & 0.228 & 0.350 & 0.427 & 0.122 & -0.277 & 0.006 & 0.012 &-0.005 \\
           64a48.5& 2.0 &82.5\degr&  5.44 & 0.144 & 0.220 & 0.348 & 0.437 & 0.128 & -0.306 & 0.002 & 0.007 &-0.007 \\
           64a49 &  2.0 & 90\degr &  5.43 & 0.145 & 0.200 & 0.361 & 0.444 & 0.161 & -0.329 & 0.000 &-0.000 &-0.005 \\
           \hline
         \end{array}
     $$ 
   \end{table*}
}

\subsection{The $\Lambda$-effect from the minimal tau-approximation}

To have some understanding of the numerical results, we
compare with the simplistic tau-approximation (hereafter MTA)
in real space (e.g.\ Blackman \& Field \cite{BlackField2002};
Brandenburg et al.\ \cite{Brandenburgea2004}).
Unlike the usual first-order smoothing approximation where nonlinearities
in the fluctuations are neglected, they are retained in an approximate
manner in MTA.

In order to develop a theory for the Reynolds stress,
$Q_{ij} = \overline{u_i u_j}$,
we derive an equation for its time derivative,
\EQ
\dot{Q}_{ij} = \overline{\dot{u}_i u_j} + \overline{u_i \dot{u}_j}\;.
\label{dotQij}
\EE
In the absence of large-scale flows, i.e. $\meanv{U} = 0$, the
equation for the fluctuating part can be written as
\EQ
\dot{u}_i = N_i - 2\, \varepsilon_{imn}\Omega_m u_n + f_i\;,
\label{duidt}
\EE
where $N_i=-u_k \pd_k u_i - \cst \pd_i\ln \rho$ is a nonlinear term.
Multiplying Eq.~(\ref{duidt}) by $u_j$ gives
\EQ
u_j \dot{u}_i = u_j N_i - 2\, u_j \varepsilon_{imn}\Omega_m u_n + u_j f_i\;.
\EE
Inserting this into Eq.~(\ref{dotQij}) yields
\EQA
\dot{Q}_{ij} = -2\,\epsilon_{jkl}\Omega_k Q_{il} -2\,\epsilon_{ikl}\Omega_k Q_{jl} + \overline{u_i f_j} + \overline{u_j f_i} + T_{ij}\;, \label{equ:mtastress}
\EEA
where $T_{ij}=\overline{u_i N_j}+\overline{u_j N_i}$
are the triple correlations.
Under the assumption of periodic boundary conditions, this term
can be written in the form
\EQ
T_{ij} = -\cst (\overline{u_i \pd_j \ln \rho} + \overline{u_j \pd_i \ln \rho}) + \overline{u_i u_j \DIV{\ve{u}}}\;. \label{equ:triple}
\EE
In MTA the higher than second-order terms, i.e.\ $T_{ij}$, in the
equations of turbulent correlations are retained in
a collective manner by parametrizing them with a term that is equal
to the original correlation divided by a relaxation time, i.e.\
\EQ%
T_{ij} = -\tau^{-1} Q_{ij}\;.\label{equ:mtaapprox}%
\EE%
Now the equations
can be solved for the stresses in terms of the forcing. A simple
ansatz for parameterizing the forcing is given in terms of the
non-rotating equilibrium solution,%
\EQ%
\overline{u_i f_j} + \overline{u_j f_i} = \tau^{-1}
Q_{ij}^{(0)}\;.\label{equ:relax}%
\EE%
Inserting the parameterizations (\ref{equ:mtaapprox}) and
(\ref{equ:relax}) into Eq.~(\ref{equ:mtastress}) yields%
\EQA%
\dot{Q}_{ij} = -2\,\epsilon_{jkl}\Omega_k Q_{il}
-2\,\epsilon_{ikl}\Omega_k Q_{jl} - \frac{1}{\tau} (Q_{ij} -
Q_{ij}^{(0)})\;;\label{equ:MTAmodel}%
\EEA%
see Appendix~\ref{app:mtaeqs} for more details on the equations used
in the MTA-model. The model equations are very similar to those of
Ogilvie (\cite{Ogilvie2003}; see also Garaud \& Ogilvie
\cite{GarOgil2005}) who studied hydrodynamic and magnetohydrodynamic
turbulence and angular momentum transport due to shear flows and
the magnetorotational instability.
Among other things, they also introduced an isotropization term that
causes decaying turbulence to become isotropic.
Some of our models give explicit evidence of such a term.

\begin{figure}[t]
\centering
\includegraphics[width=0.45\textwidth]{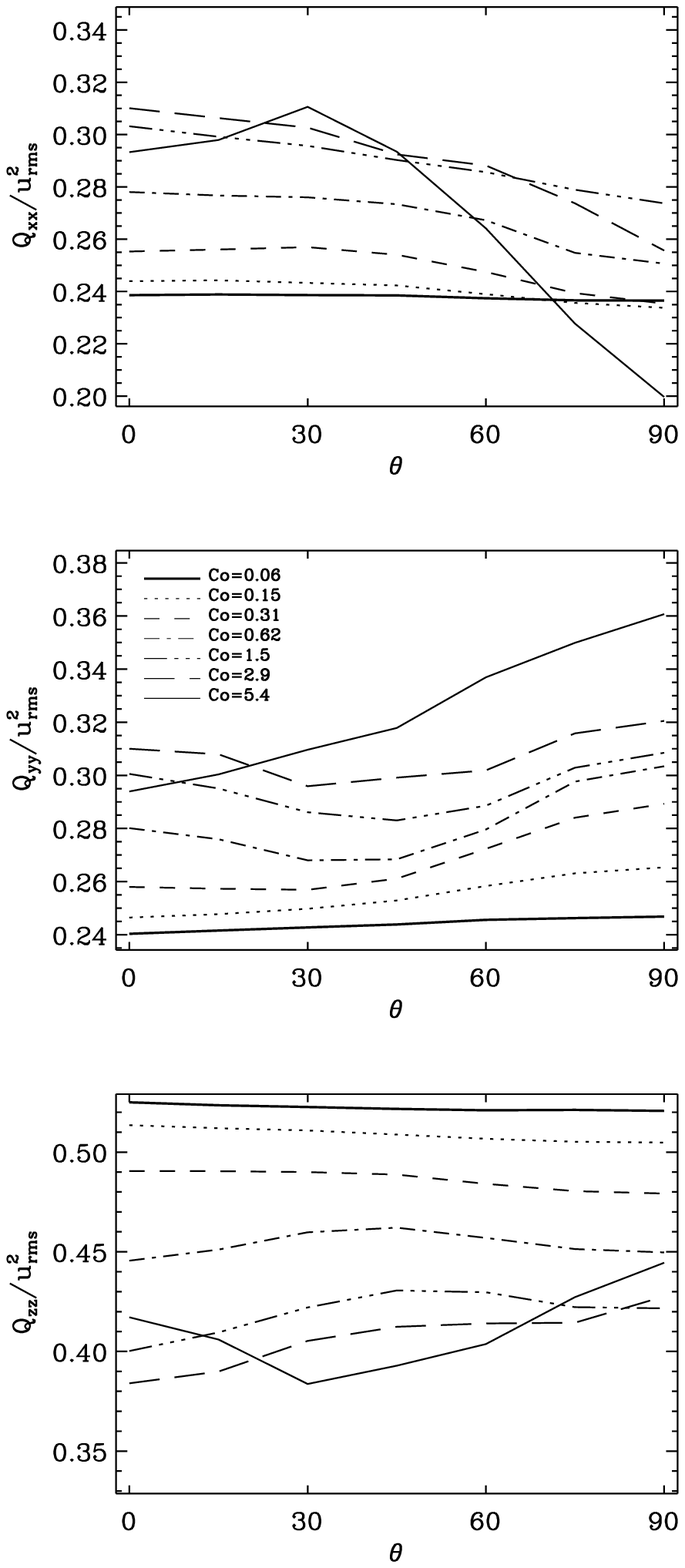}
\caption{Volume-averaged Reynolds stress components $Q_{xx}$ (top),
  $Q_{yy}$ (middle), and $Q_{zz}$ (bottom), normalized by the square
  of the rms-velocity, as functions of latitude and rotation from the
  turbulence simulations listed in Table~\ref{tab:Re12}. Coriolis
  number, as defined in Eq.~(\ref{equ:Coriolis}), varies as indicated
  in the legend in the middle panel.}
\label{fig:diag}
\end{figure}

\begin{figure}[h]
\centering
\includegraphics[width=0.45\textwidth]{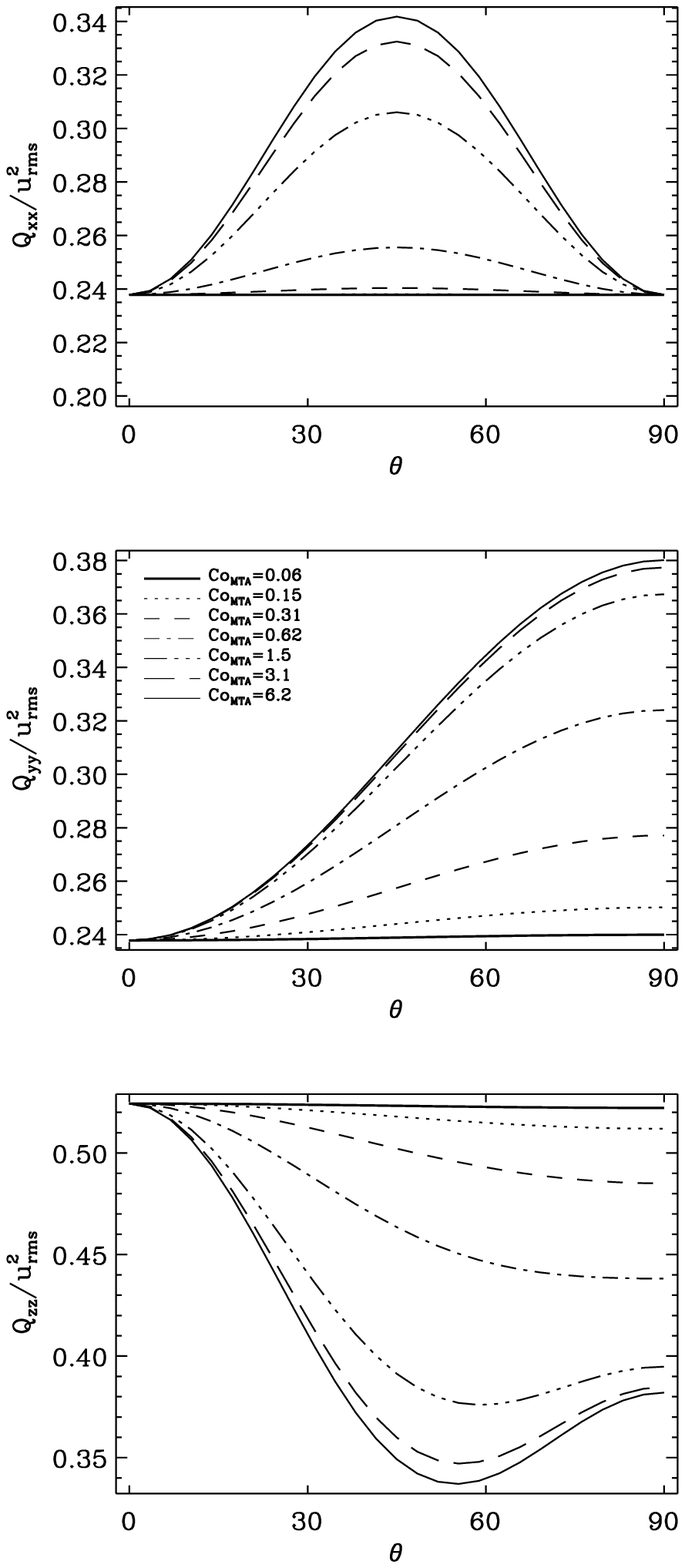}
\caption{Diagonal Reynolds stress components from the MTA-model. Here
  the Coriolis number is defined via Eq.~(\ref{equ:CoMTA}) with $\St =
  1$ and $\xi=0$.}
\label{fig:diag_mta}
\end{figure}

\begin{figure}[h]
\centering
\includegraphics[width=0.45\textwidth]{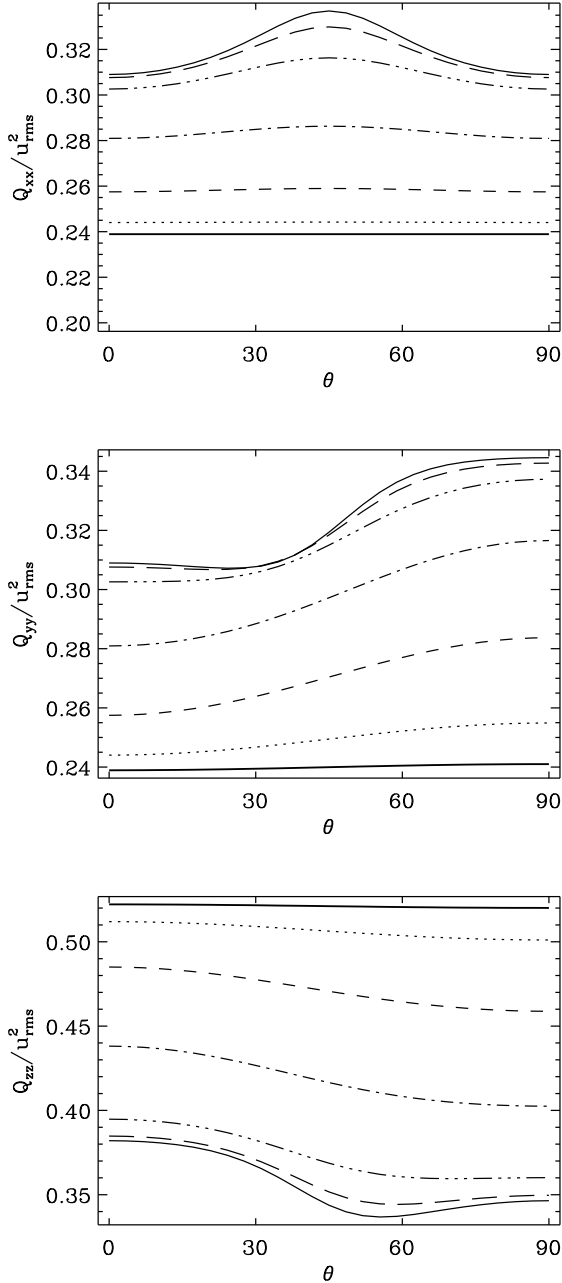}
\caption{Same as Fig.~\ref{fig:diag_mta} but with $\St = \xi =
  1$. Compare with Figs.~\ref{fig:diag} and \ref{fig:diag_mta}.}
\label{fig:diag_mta_St1_rotq}
\end{figure}

The rotational influence in the MTA-model is measured by the Coriolis
number%
\EQ%
\Co_{\rm MTA} = \frac{2\,\Omega_0\,\St}{\urms \kef}
\equiv\Co\,\St\;, %\label{equ:CoMTA}
\label{equ:CoMTA}%
\EE%
where%
\EQ%
\St = \tau \urms \kef \label{equ:strouhal}%
\EE%
is the Strouhal number, which is the main free parameter in the
model.  

On account of previous results on similar systems of forced
turbulence (e.g.\ Brandenburg \& Subramanian \cite{BranSubra2005})
$\St = 1$ is used as our reference model.
To reproduce the numerical results, we introduce
an empirical isotropization term in the MTA-model; see
Sect.~\ref{sec:diaresults} and in particular Eqs.~(\ref{equ:rotiso})
and (\ref{equ:rotiso2}). The use of this term is regulated by 
another free parameter $\xi$, which can obtain the values zero or 
unity.

The MTA-model results were obtained by advancing the time-dependent 
equations (\ref{equ:mtastress}) with the parameterizations
(\ref{equ:mtaapprox}) and (\ref{equ:relax}) until a stationary
solution was reached.

\section{Results}
\label{sec:results}

\subsection{Diagonal components of the stress}
\label{sec:diaresults}
Consider turbulence where the intensity in the vertical direction is
stronger than the intensities in the horizontal directions. This
situation is encountered in the turbulence caused by the convective
instability. First we consider the case with the maximum anisotropy
that was achieved with the present model. Table \ref{tab:Re12}
summarizes the results for seven values of the Coriolis numbers
ranging from 0.06 to roughly 5.4. Calculations at seven latitudes from
the pole ($\theta=0\degr$) to the equator ($\theta=90\degr$) with
equidistant intervals of $\Delta \theta = 15\degr$ were made with each
Coriolis number, ${\rm Re} \approx 12 \ldots 14$ in all runs. In this
set of runs, $f_0 = 10^{-6}$ and $f_1 = 0.2$ were used in
Eq.~(\ref{equ:forceaniso}). 

Figure~\ref{fig:diag} shows the diagonal components of the Reynolds
tensor as functions of latitude and Coriolis number from the numerical
turbulence simulations. As rotation is increased, the magnitudes of the
horizontal components $\qxx$ and $\qyy$ increase monotonically
while $\qzz$ decreases, which illustrates an isotropizing effect of
rotation on the turbulence.
In the following, we refer to this effect as the rotational isotropization
of turbulence.
This effect is a purely empirical and refers to the observation that
increased rotation leads to stronger mixing which washes out
anisotropies that are caused by other effects.
Of course, rotation itself can cause the turbulence to become
anisotropic, but this seems to play a role only at much higher
rotation rates.
This is indeed seen 
in the most rapidly rotating case where the behavior is more
complex. The MTA-model, Fig.~\ref{fig:diag_mta}, on the other hand,
fails to reproduce isotropization of turbulence when rotation is
included. The behavior is most obvious at the pole where rotation
contributes no net effect to linear order, see
Appendix~\ref{app:mtaeqs}, Eqs.~(\ref{equ:appqxx}) to
(\ref{equ:appqzz}). The same is true for the $\qxx$ component at the
equator.

Some persistent trends arise as a function of latitude in the
numerical simulations; for instance, $\qxx$ peaks at the pole and decreases toward
the equator except again for the fastest rotation. An approximately
opposite trend is seen for $\qyy$ for slow and rapid rotation, whereas
in the intermediate range a minimum appears at mid-latitudes. For slow
rotation, $\qzz$ behaves approximately in the opposite
way to $\qyy$, having a maximum at the pole and a minimum at
the equator, whereas for rapid rotation the trend is reversed. There
is also a persistent maximum at mid-latitudes. The MTA-model, however,
fails to reproduce most of the characteristics of the latitude
distribution of the diagonal stresses. This indicates that nonlinear
terms, which are manifestly not described adequately well by the
MTA-assumption, are the deciding factor in determining
the behavior of the diagonal stresses.

To capture the rotational
isotropization with the MTA-model at least qualitatively, we experimented by adding a term
\begin{equation}
\dot{Q}_{ij} = \ldots - \frac{\xi}{\tau} F_{\rm rot} (\qij - \onethird \delta_{ij}Q)\;, \label{equ:rotiso}
\end{equation}
on the rhs of Eq.~(\ref{equ:MTAmodel}). Here $Q$ is the trace of
$\qij$ and
\begin{equation}
F_{\rm rot} = \frac{3\,\Co^2_{\rm MTA}}{1+\Co_{\rm MTA}^2}\;.\label{equ:rotiso2}
\end{equation}
The functional form of $F_{\rm rot}$ is chosen purely empirically so that
the magnitudes of the off-diagonal stresses in comparison to the
simulations are fairly accurately reproduced with the MTA-model (see
Fig.~\ref{fig:lambda_mta_St1_rotq}).

Figure~\ref{fig:diag_mta_St1_rotq} shows the results for the diagonal
components with $\St=\xi=1$. Now the magnitudes of the
turbulence intensities are more in line with the full numerical
simulations, although the latitude distribution is still manifestly
wrong. Although the off-diagonal components are much better
represented by the linear terms appearing in the equation of the
Reynolds stress (see Sect.~\ref{subsec:offdia}), the rotational
isotropization term helps for reducing their magnitudes closer to the
levels seen in the direct simulations also in that case.

The functional form of $F_{\rm rot}$ in
Eq.~(\ref{equ:rotiso}) indicates that  $F_{\rm rot}
\rightarrow 3$ for rapid rotation. This behaviour cannot be justified based on the
present numerical data (see Fig.~\ref{fig:diag}).

\subsection{Anisotropy of the turbulence}
Turbulence anisotropies can be characterized by the quantities
\begin{eqnarray}
A_{\rm H} & = & \frac{\qyy - \qxx}{\urms^2}\;, \label{equ:AH} \\
A_{\rm V} & = & \frac{\qxx + \qyy - 2\,\qzz}{\urms^2}\;. \label{equ:AV}
\end{eqnarray}
The importance of these quantities is that, for slow rotation, they can
be considered as proxies of the $\Lambda$-effect according to
$\Lambda_{\rm H} \approx 2\,\tauc A_{\rm H}$ and $\Lambda_{\rm V}
\approx 2\, \tauc A_{\rm V}$ (e.g.\ R\"udiger
\cite{Ruediger1980}, \cite{Ruediger1989}), 
where $\tauc$ is the correlation time of the turbulence.

Table \ref{tab:Re12} shows that for slow rotation $A_{\rm H}$
increases monotonically from the pole to the equator. For $\Co > 0.3$,
$A_{\rm H}$ exhibits a negative minimum at mid-latitudes and reaches a
positive maximum at the equator. The maximum at the equator can be
explained as resulting from the surviving ($y$-)component of the
Coriolis force at low latitudes. The relation between the horizontal
$\Lambda$-effect and the anisotropy parameter $A_{\rm H}$ is at best
poorly confirmed by the numerical results. It must also be noted that
$A_{\rm H}$ is smaller by at least an order of magnitude in comparison
to $A_{\rm V}$, which also enters the equation for $\qxy$ but with a
higher order in $\Co$ (see Eq.~\ref{equ:mtamodelqxy}).

The vertical anisotropy, $A_{\rm V}$, on the other hand, retains its
sign in all models. For slow rotation the absolute value of $A_{\rm
  V}$ decreases monotonically from the pole toward the equator. The
latitude distribution for rapid rotation is approximately opposite,
although an additional minimum ($\Co \approx 0.6\ldots1.5$) or a maximum
($\Co \approx 5.4$) can occur at mid-latitudes. Here the correspondence
between $A_{\rm V}$ and $\Lambda_{\rm V}$ holds at least for the sign
for all calculations, bar the few cases in the intermediate rotation
regime where $\qyz$ is positive (see below).

\subsection{Off-diagonal components of the stress}
\label{subsec:offdia}
\begin{figure}[t]
\centering
\includegraphics[width=0.45\textwidth]{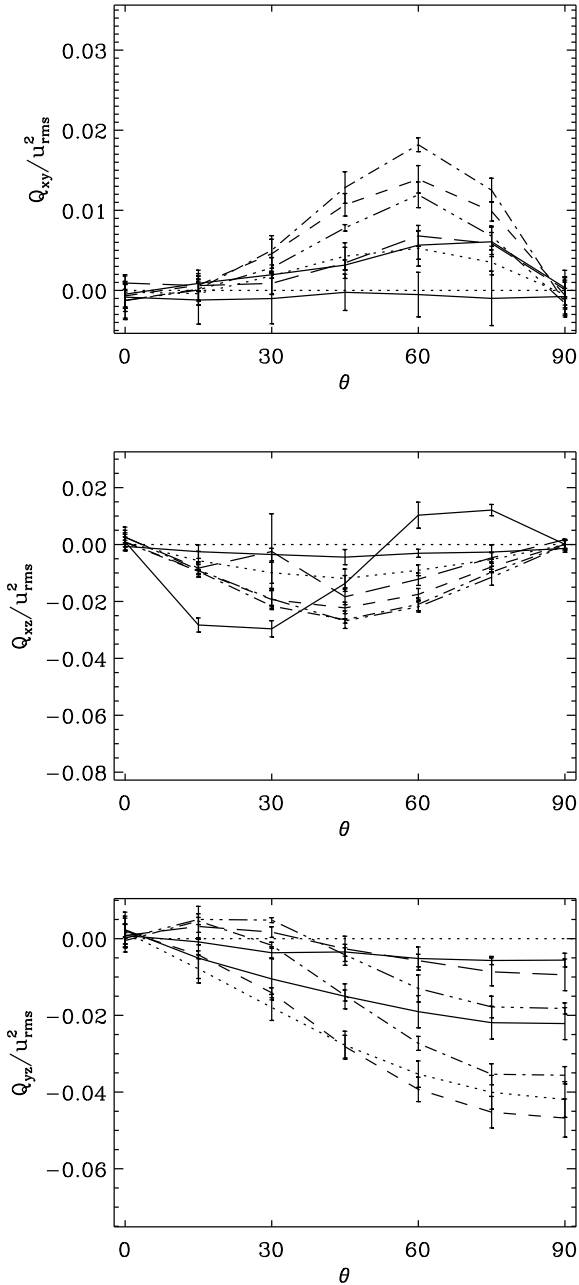}
\caption{Same as Fig.~\ref{fig:diag} but for the off-diagonal stress
  components $Q_{xy}$ (top), $Q_{xz}$ (middle), and $Q_{yz}$
  (bottom). Linestyles as in Fig.~\ref{fig:diag}}
\label{fig:lambda}
\end{figure}

In stellar convection zones, the off-diagonal Reynolds stresses
contribute to the angular momentum transport and work to generate
($\Lambda$-effect) or to smooth out (turbulent viscosity) differential
rotation. In the present case only the former effect is in
operation. Figure~\ref{fig:lambda} summarizes the results for the runs
listed in Table~\ref{tab:Re12}.

\begin{figure}[h]
\centering
\includegraphics[width=0.45\textwidth]{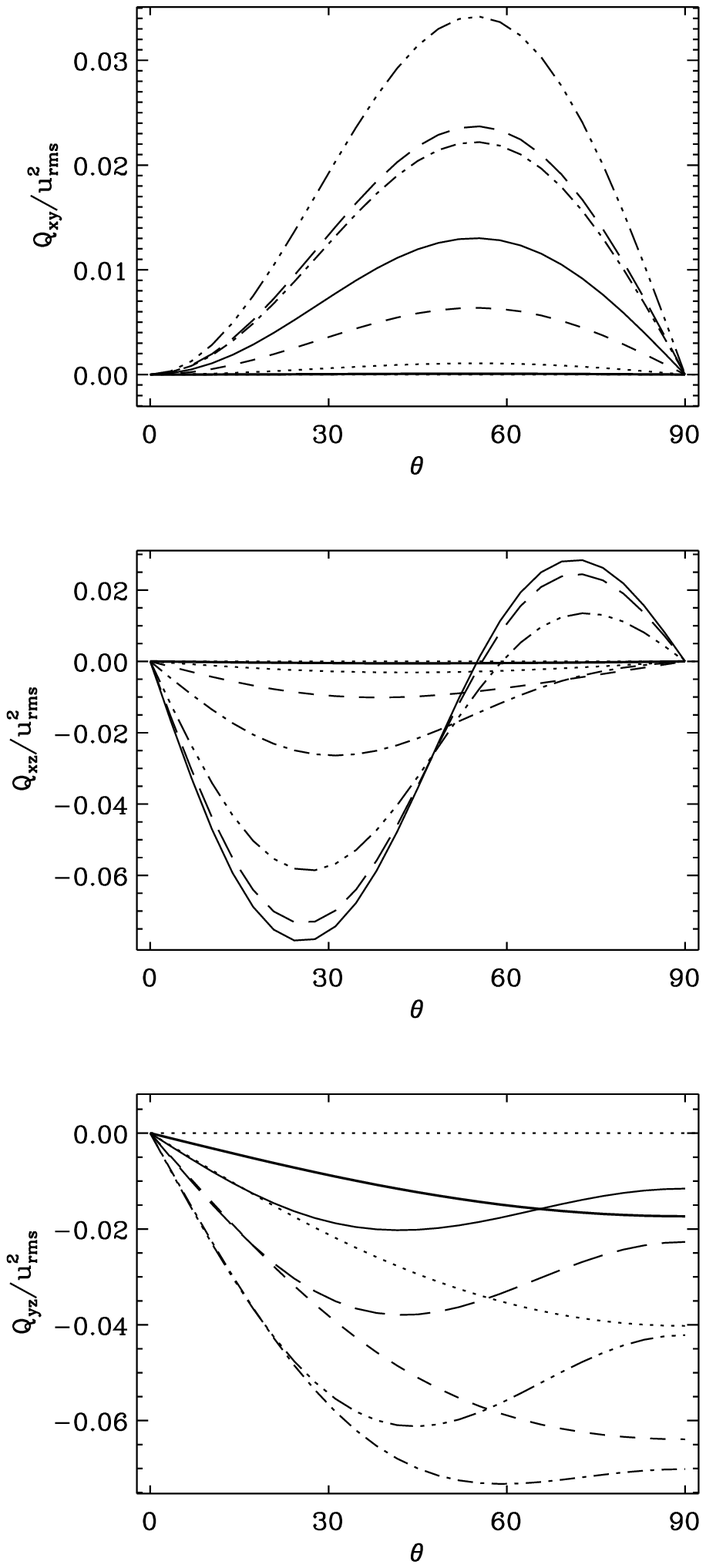}
\caption{Same as Fig.~\ref{fig:diag_mta} but for the off-diagonal
  stress components $Q_{xy}$ (top), $Q_{xz}$ (middle), and $Q_{yz}$
  (bottom). ${\rm St} = 1$, $\xi = 0$. Linestyles as in
  Fig.~\ref{fig:diag_mta}}
\label{fig:lambda_mta_St1}
\end{figure}

\begin{figure}[h]
\centering
\includegraphics[width=0.45\textwidth]{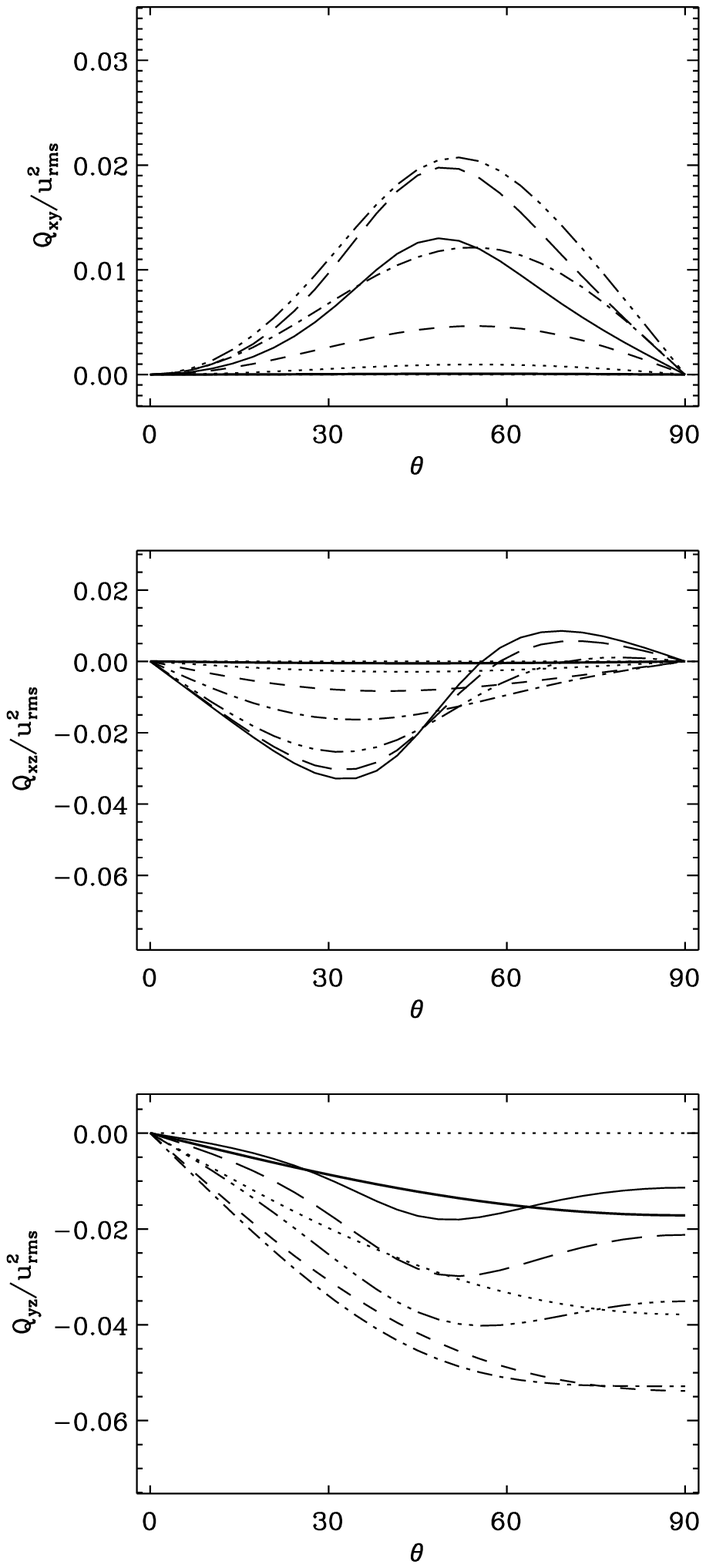}
\caption{Same as Fig.~\ref{fig:lambda_mta_St1} but with ${\rm
    St} = \xi = 1$. Linestyles as in Fig.~\ref{fig:diag_mta}.}
\label{fig:lambda_mta_St1_rotq}
\end{figure}

\begin{figure}[h]
\centering
\includegraphics[width=0.45\textwidth]{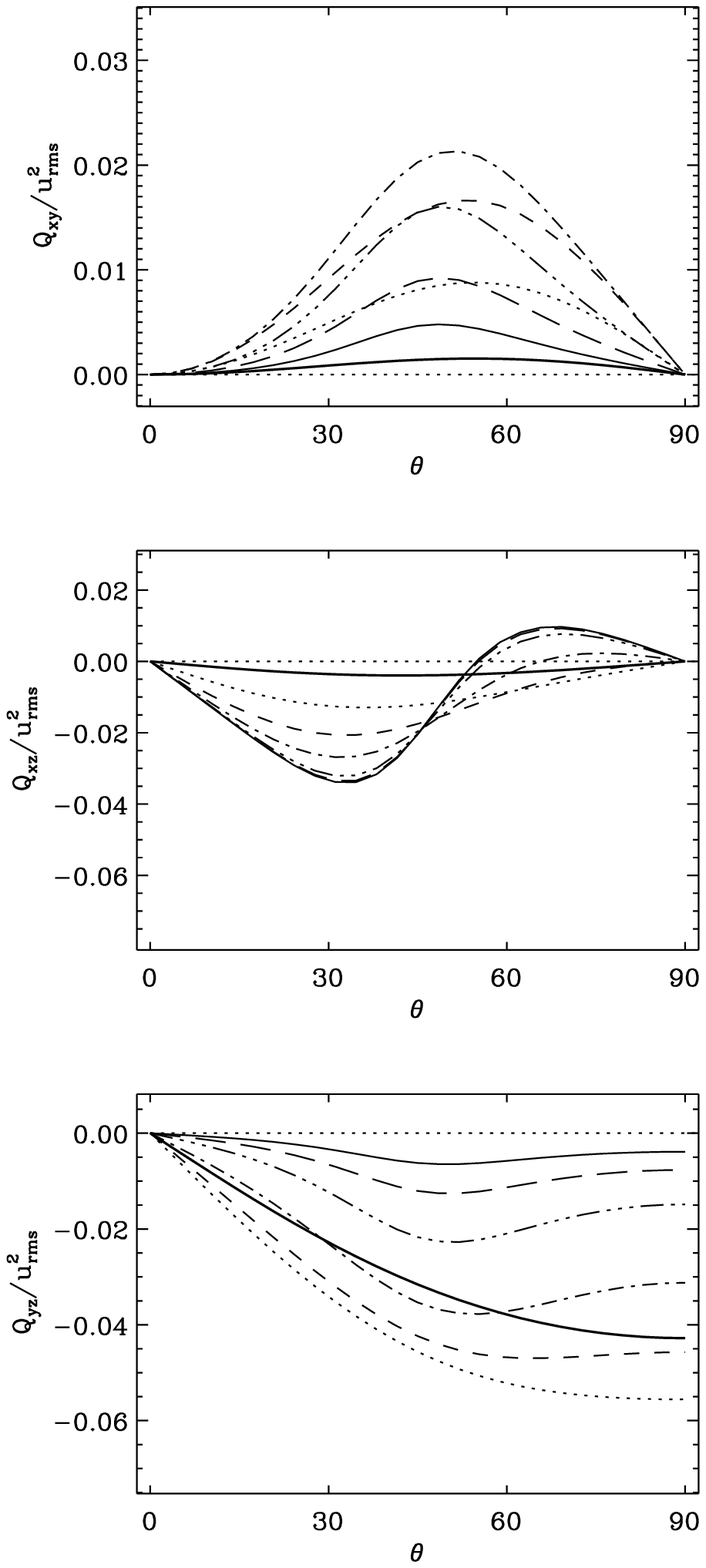}
\caption{Same as Fig.~\ref{fig:lambda_mta_St1} but with ${\rm
    St} = 2$, $\xi = 1$. Linestyles as in Fig.~\ref{fig:diag_mta}.}
\label{fig:lambda_mta_St2_rotq}
\end{figure}

The $\qxy$ component of the stress corresponds to $Q_{\theta \phi}$ in
spherical coordinates and is responsible for latitudinal angular
momentum transport. In the simulations this component is always
positive and the latitude distribution peaks at latitude $30\degr$
(see the uppermost panel of Fig.~\ref{fig:lambda}). The sign is in
accordance with solar observations (Ward \cite{Ward1965}; Pulkkinen \&
Tuominen \cite{PulkTuo1998}) and analytical turbulence models
(Kitchatinov \& R\"udiger \cite{KitRued1993}, \cite{KitRued2005}). The
latitude distribution in the rapid rotation regime is significantly
different from convection simulations where $\qxy$ is 
sharply concentrated near the equator (Chan \cite{Chan2001};
K\"apyl\"a et al.\ \cite{Kaepylaeea2004}; Hupfer et al.\
\cite{Hupferea2005}; R\"udiger et al.\ \cite{Ruedigerea2005a}). The
reason for this difference is still unclear but is evidently
related to the physics that have been omitted in the present
study. The MTA-model captures the latitude dependence rather well, but
the magnitude of the stress is clearly too great (see
Fig.~\ref{fig:lambda_mta_St1}). If the rotational isotropization term,
Eq.~(\ref{equ:rotiso}), is taken into account (see the uppermost panel
of Fig.~\ref{fig:lambda_mta_St1_rotq}), the agreement is also better
for the magnitude. Keeping the forcing and rotation rate $\Omega_0$
fixed, the best agreement with the 3D models is found if $\St=2$ is used,
see Fig.~\ref{fig:lambda_mta_St2_rotq}. 

While the analogue of the $\qxz$ component does not play a direct
role in the angular momentum balance in stars, it can still contribute
via generating meridional flows (e.g.\ R\"udiger
\cite{Ruediger1989}). In the numerical simulations we find that, for
all cases except the most rapidly rotating one, $\qxz$ is negative and
peaks at $\theta=45\degr$ (middle panel of Fig.~\ref{fig:lambda}). For
$\Co \approx 5.4$, however, the sign changes near the equator, with
positive values toward the equator and negative ones toward the
pole. This is at odds with the analytical result of R\"udiger et al.\
(\cite{Ruedigerea2005b}), but does agree with the results of
convection simulations (Pulkkinen et al.\ \cite{Pulkkinenea1993};
K\"apyl\"a et al.\ \cite{Kaepylaeea2004}).

The MTA-model gives qualitatively similar results, although the sign
change occurs at significantly slower rotation; see the middle panel
of Fig.~\ref{fig:lambda_mta_St1}. Rotational isotropization helps to
correct the magnitude, but not the earlier occurrence of the sign
change (Figs.~\ref{fig:lambda_mta_St1_rotq} and
\ref{fig:lambda_mta_St2_rotq}).

\begin{figure*}%[h]
\centering
\includegraphics[width=1.\textwidth]{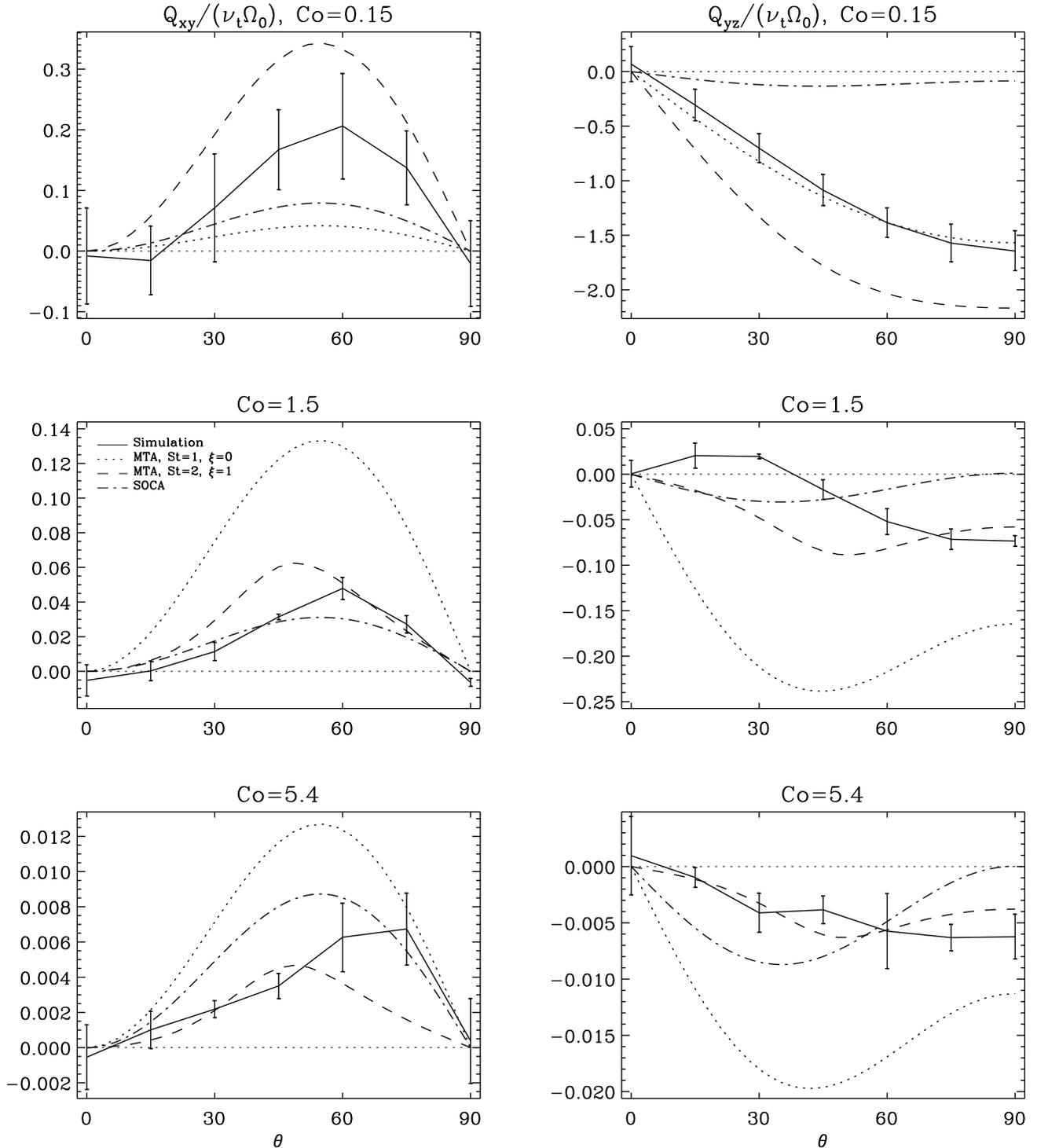}
\caption{Comparison of the numerical simulations (solid lines), two
  MTA-models with $\St=1$, $\xi=0$ (dotted), and $\St=2$, $\xi=1$
  (dashed), with the SOCA results of KR05 (dash-dotted). Left (right)
  hand panels show $\qxy/(\nut \Omega_0)$ ($\qyz/(\nut \Omega_0)$) for
  Coriolis numbers 0.15 (uppermost panels), 1.5 (middle), and 5.4
  (lower panels).}
\label{fig:pcomp}
\end{figure*}

Since the turbulent intensity of the vertical ($z$-)motions is greater
than the horizontal ones, the expectation is that the stress component
$\qyz$ is negative, i.e. that $\Lambda_{\rm V} \propto A_{\rm V}$
(Biermann \cite{Biermann1951}). This is indeed seen in the simulations
quite consistently (lowermost panel of Fig.~\ref{fig:lambda}),
although at intermediate rotation low positive values can occur at high latitudes. The highest values of $\qyz$ occur for $\Co \approx 0.3$ as
opposed to $\Co \approx 0.6$ for $\qxy$. Rotational quenching of
$\qyz$ seems to be stronger and occur for lower $\Co$ than for $\qxy$
(see also Fig.~\ref{fig:pcomp}). A similar trend was seen in convection
simulations by K\"apyl\"a et al.\ (\cite{Kaepylaeea2004}). The
consistently positive values of $\qyz$ for rapid rotation seen in
convection simulations (K\"apyl\"a et al.\
\cite{Kaepylaeea2004}; Chan 2007, private communication) do not occur
in the present calculations.

The value of $\qyz$ always reaches a maximum at the equator. This
contradicts with analytical results derived under first-order
smoothing (Kitchatinov \& R\"udiger \cite{KitRued1993},
\cite{KitRued2005}) and the MTA-model that predicts a maximum around
$\theta=45\degr$ for intermediate and rapid rotation. The rotational
isotropization term is again needed to reduce the magnitude
of the stress. It seems that to reproduce $\qyz$ correctly,
one would need to apply somewhat stronger isotropization than what is
presently used (see Figs.~\ref{fig:lambda_mta_St1_rotq} and
\ref{fig:lambda_mta_St2_rotq}).

One conclusion that can be drawn from the MTA-model results is that,
although the diagonal Reynolds stresses are quite poorly reproduced in
comparison to the 3D simulations, the off-diagonals have most of the
qualitative features correct. This seems to imply that, to model the
off-diagonals, the exact parameterization of the nonlinearities in the
equation of the Reynolds stress is not crucial. The
empirical rotational isotropization term helps to capture some of the
missing features for the diagonal components and reduces the
magnitudes of the off-diagonals to the level that is also seen in the
simulations.

\begin{figure*}%[h]
\centering
\includegraphics[width=1.\textwidth]{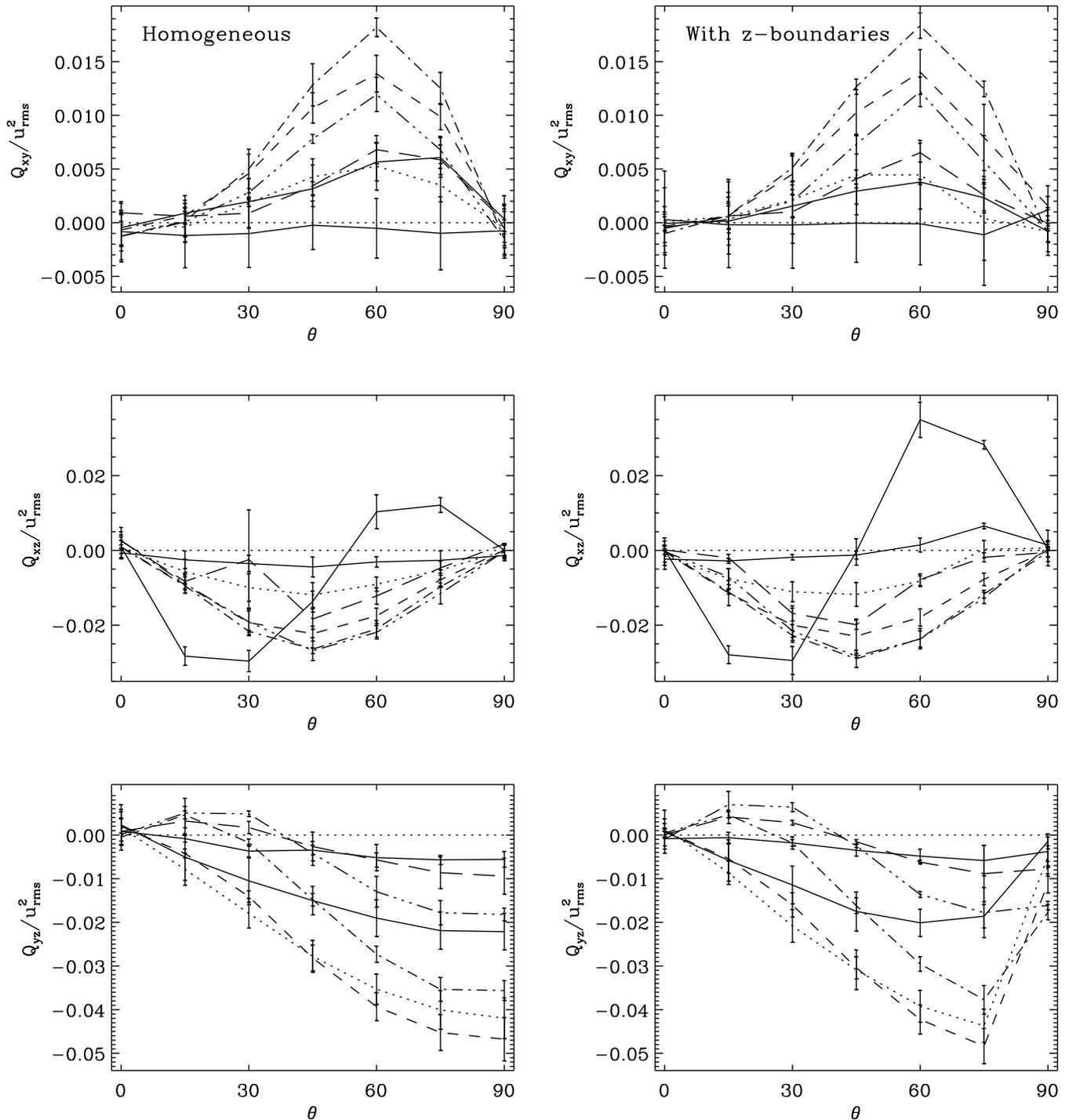}
\caption{Same as Fig.~\ref{fig:lambda} but for homogeneous (left
  panel) and inhomogeneous (right) simulations.}
\label{fig:lambda_inhom}
\end{figure*}

\onltab{2}{
   \begin{table*}
   \centering
   \caption[]{Summary of the turbulence anisotropies and normalized Reynolds stresses,  $\tilde{Q}_{ij} = Q_{ij}/\urms^2$, for a set of runs with varying turbulence anisotropy. $\Omega_0 = 0.1$, $\kef/k_1=5$, $\nu = 2 \cdot 10^{-3}$, and grid resolution $64^3$ was used in all runs.}
      \label{tab:varaniso}
%      \vspace{-0.5cm}
     $$
         \begin{array}{p{0.055\linewidth}cccccccrcrrr}
           \hline
           \noalign{\smallskip}
           Run      & \Omega_0  & \theta & \Co & \urms & \tilde{Q}_{xx} & \tilde{Q}_{yy} & \tilde{Q}_{zz} & A_{\rm H} & A_{\rm V} & \tilde{Q}_{xy} & \tilde{Q}_{xz} & \tilde{Q}_{yz} \\
           \noalign{\smallskip}
           \hline
           \noalign{\smallskip}
           64a15 & 0.1 &  0\degr & 0.31 & 0.126 & 0.255 & 0.258 & 0.490 & 0.003 & -0.468 &-0.000 & 0.002 & 0.002 \\
           64a16 & 0.1 & 15\degr & 0.31 & 0.127 & 0.256 & 0.257 & 0.490 & 0.001 & -0.468 & 0.001 &-0.009 &-0.004 \\
           64a17 & 0.1 & 30\degr & 0.31 & 0.127 & 0.257 & 0.257 & 0.490 &-0.000 & -0.466 & 0.005 &-0.019 &-0.014 \\
           64a18 & 0.1 & 45\degr & 0.31 & 0.127 & 0.254 & 0.261 & 0.489 & 0.007 & -0.462 & 0.011 &-0.022 &-0.028 \\
           64a19 & 0.1 & 60\degr & 0.31 & 0.127 & 0.247 & 0.272 & 0.484 & 0.025 & -0.448 & 0.014 &-0.017 &-0.039 \\
           64a20 & 0.1 & 75\degr & 0.31 & 0.127 & 0.239 & 0.284 & 0.480 & 0.045 & -0.437 & 0.010 &-0.008 &-0.045 \\
           64a21 & 0.1 & 90\degr & 0.31 & 0.127 & 0.235 & 0.289 & 0.479 & 0.054 & -0.434 &-0.000 & 0.000 &-0.047 \\
           \hline
           \noalign{\smallskip}
           64c15 & 0.1 &  0\degr & 0.34 & 0.115 & 0.275 & 0.275 & 0.453 & 0.000 & -0.357 & 0.001 & 0.003 & 0.000 \\
           64c16 & 0.1 & 15\degr & 0.34 & 0.116 & 0.275 & 0.276 & 0.454 & 0.001 & -0.357 & 0.003 &-0.009 &-0.007 \\
           64c17 & 0.1 & 30\degr & 0.34 & 0.116 & 0.277 & 0.275 & 0.452 &-0.002 & -0.353 & 0.006 &-0.017 &-0.018 \\
           64c18 & 0.1 & 45\degr & 0.34 & 0.116 & 0.275 & 0.280 & 0.449 & 0.006 & -0.343 & 0.011 &-0.019 &-0.030 \\
           64c19 & 0.1 & 60\degr & 0.34 & 0.115 & 0.271 & 0.290 & 0.443 & 0.019 & -0.324 & 0.014 &-0.015 &-0.038 \\
           64c20 & 0.1 & 75\degr & 0.34 & 0.116 & 0.265 & 0.300 & 0.439 & 0.035 & -0.314 & 0.010 &-0.007 &-0.043 \\
           64c21 & 0.1 & 90\degr & 0.34 & 0.116 & 0.262 & 0.304 & 0.439 & 0.042 & -0.311 & 0.002 & 0.001 &-0.044 \\
           \hline
           \noalign{\smallskip}
           64b15 & 0.1 &  0\degr & 0.33 & 0.119 & 0.299 & 0.299 & 0.406 & 0.000 & -0.215 & 0.001 & 0.001 & 0.002 \\
           64b16 & 0.1 & 15\degr & 0.33 & 0.119 & 0.300 & 0.299 & 0.405 &-0.000 & -0.211 & 0.002 &-0.005 &-0.005 \\
           64b17 & 0.1 & 30\degr & 0.33 & 0.119 & 0.302 & 0.299 & 0.404 &-0.003 & -0.207 & 0.004 &-0.011 &-0.012 \\
           64b18 & 0.1 & 45\degr & 0.33 & 0.119 & 0.301 & 0.303 & 0.400 & 0.002 & -0.195 & 0.007 &-0.012 &-0.019 \\
           64b19 & 0.1 & 60\degr & 0.33 & 0.119 & 0.300 & 0.308 & 0.396 & 0.008 & -0.185 & 0.010 &-0.011 &-0.024 \\
           64b20 & 0.1 & 75\degr & 0.33 & 0.119 & 0.297 & 0.313 & 0.395 & 0.016 & -0.179 & 0.006 &-0.004 &-0.026 \\
           64b21 & 0.1 & 90\degr & 0.33 & 0.119 & 0.295 & 0.317 & 0.393 & 0.022 & -0.175 & 0.003 &-0.001 &-0.027 \\
           \hline
           \noalign{\smallskip}
           64d15 & 0.1 &  0\degr & 0.34 & 0.115 & 0.321 & 0.318 & 0.365 &-0.002 & -0.091 & 0.001 & 0.000 & 0.001 \\
           64d16 & 0.1 & 15\degr & 0.34 & 0.115 & 0.320 & 0.320 & 0.364 & 0.000 & -0.087 & 0.001 &-0.004 &-0.002 \\
           64d17 & 0.1 & 30\degr & 0.34 & 0.115 & 0.321 & 0.320 & 0.363 &-0.001 & -0.086 & 0.001 &-0.006 &-0.005 \\
           64d18 & 0.1 & 45\degr & 0.34 & 0.115 & 0.321 & 0.323 & 0.360 & 0.002 & -0.076 & 0.003 &-0.006 &-0.009 \\
           64d19 & 0.1 & 60\degr & 0.34 & 0.115 & 0.324 & 0.323 & 0.357 &-0.001 & -0.066 & 0.004 &-0.006 &-0.012 \\
           64d20 & 0.1 & 75\degr & 0.34 & 0.115 & 0.323 & 0.325 & 0.356 & 0.002 & -0.065 & 0.003 &-0.004 &-0.013 \\
           64d21 & 0.1 & 90\degr & 0.34 & 0.115 & 0.319 & 0.327 & 0.357 & 0.009 & -0.069 & 0.000 &-0.000 &-0.013 \\
           \hline
         \end{array}
     $$ 
   \end{table*}
}

\subsubsection{Comparison with inhomogeneous simulations}
The homogeneous setup used so far prevents any mean flows from being
generated. This is good for the purpose of testing the sole effect of 
turbulent velocity field on the Reynolds stresses, but can be argued to 
be unphysical because astronomical objects where turbulence is 
important; e.g., the solar convection zone has boundaries and cannot be 
considered homogeneous.

To test how much the assumption of homogeneity affects the results,
we made a set of simulations with a setup where the $z$-boundaries are 
impenetrable. For the horizontal velocity components, stress-free boundary 
conditions are used, i.e.\
\begin{equation}
U_{x,z} = U_{y,z} = U_z = 0\;.
\end{equation}
To compare with the homogeneous simulations and at the same time 
minimize the effects of the boundaries, we average the results over 
$-\threefourths \pi < z < \threefourths \pi$ and the full $(x,y)$-extents.

Mean flows 
are generated in the runs that, however, are small in comparison to the
rms-velocity of the turbulence. The exception to this trend is the 
equator, i.e.\ 
$\theta = 90\degr$, where a large mean shear flow develops, for which
\begin{equation}
\frac{\Delta U_y}{\Delta z} \approx -\Omega_0\;.
\end{equation}
Similar flows have been seen in convection simulations 
(Chan \cite{Chan2001}; K\"apyl\"a et al.\ \cite{Kaepylaeea2004}; 
Brandenburg \cite{Brandenburg2007}). The volume 
average of this flow is zero. For the present simulations the mean 
velocities near the $z$-boundaries can approach
the speed of sound or even exceed it. This prevents us from computing models with 
$\Omega_0 = 0.5 \ldots 2$ until a statistically saturated state.

The results for the off-diagonal stresses are compared in 
Fig.~\ref{fig:lambda_inhom}. For slow rotation, $\Co < 0.62$, the 
differences are minute at all other latitudes except at the equator. The 
tendency for $\qyz$ to have a minimum at the equator is reminiscent 
of results from convection simulations (e.g.\ Chan \cite{Chan2001}; 
K\"apyl\"a et al.\ \cite{Kaepylaeea2004}; 
R\"udiger et al.\ \cite{Ruedigerea2005b}).
For more rapid rotation, the trends for different components seem to 
diverge: $\qxy$ is somewhat reduced whereas $\qxz$ seems to increase 
somewhat and there is hardly any change for $\qyz$ apart from the 
equatorial case. For 
$\Co > 0.62$ the stresses at the equator are not saturated because
the large scale flow is not fully developed.

\subsubsection{Comparison to SOCA results}
To connect to earlier studies, the 3D simulation data and
MTA-model results are compared to the analytical SOCA results of
Kitchatinov \& R\"udiger (\cite{KitRued2005}; see also Kitchatinov \&
R\"udiger \cite{KitRued1993}). Since analytical results are only available
for the components relevant to the $\Lambda$-effect, only the
$\qxy$ and $\qyz$ components are thus considered. Furthermore, we
restrict the comparison to a subset of the models with $\Co = (0.15,
1.5, 5.4)$.

The conventional way of writing the $\Lambda$-effect is (e.g.\ R\"udiger
\cite{Ruediger1989})
\begin{eqnarray}
\qxy &=& \nut \Omega_0 H \cos \theta\;, \label{equ:conqxy} \\
\qyz &=& \nut \Omega_0 V \sin \theta\;, \label{equ:conqyz}
\end{eqnarray}
where $\nut = \onethird \urms/\kef$ is the turbulent viscosity, and
the dimensionless quantities $H$ and $V$ are given by
\begin{eqnarray}
H &=& H^{(1)} \sin^2 \theta\;, \label{equ:H} \\
V &=& V^{(0)} + V^{(1)} \cos^2 \theta\;, \label{equ:V}
\end{eqnarray}
where $H^{(1)}$, $V^{(0)}$, and $V^{(1)}=-H^{(1)}$ depend on the
Coriolis number.

The normalized stresses $Q_{ij}/(\nut \Omega_0)$ are known 
from the simulations
and the MTA-model, whereas $H$ and $V$
can be computed
analytically for the turbulence model of KR05 (see Appendix
\ref{sec:socalambda}). The results are shown in Fig.~\ref{fig:pcomp}
for three Coriolis numbers. Our Coriolis number is smaller than that
of KR05 by a factor of $2\pi$.

For the horizontal stress, the SOCA results and the rotationally
quenched MTA-model seem to fare similarly well. The former
underestimates the magnitude for small $\Co$ and overestimates it for
large $\Co$, whereas for the latter the trend is exactly the
opposite. If rotational isotropization is not taken into account, the
agreement is poor for all Coriolis numbers considered here.

For $\qyz$ the standard MTA-model is almost spot on for $\Co=0.15$,
but overestimates the magnitudes by at least a factor of two for the
other cases. As discussed in the previous section, the latitude
distribution shows a mid-latitude maximum that is not present in the
numerical simulations. When rotational isotropization is taken into
account, at least the magnitude can be reconciled with the numerical
results. The SOCA result does not fare very well in this case, 
predicting,
in general, values that are too low and an incompatible latitude distribution
with zero stress at the equator.

\begin{figure}[t]
\centering
\includegraphics[width=0.45\textwidth]{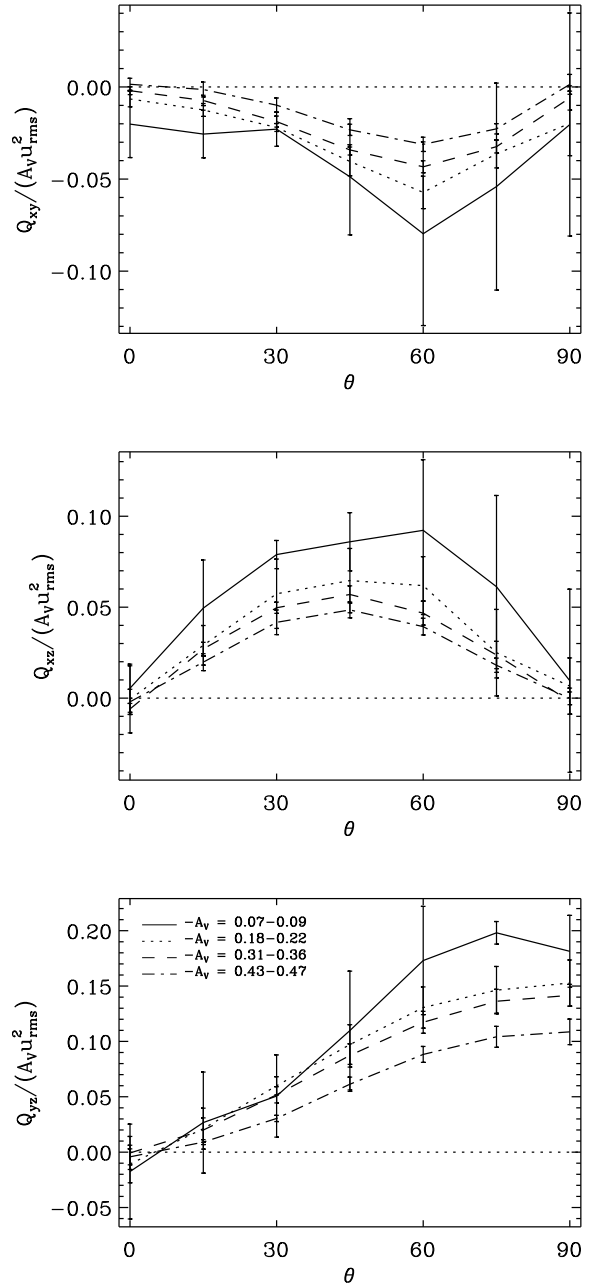}
\caption{Volume-averaged Reynolds stress components $Q_{xy}$ (top),
  $Q_{xz}$ (middle), and $Q_{yz}$ (bottom) as functions of latitude
  and vertical turbulence anisotropy $A_{\rm V}$, see
  Eq.~(\ref{equ:AV}). Line styles as indicated by the legend in the
  lower-most panel. Coriolis number in each case is roughly 0.3.}
\label{fig:lambda_varaniso}
\end{figure}

\subsection{Dependence on the amount of  anisotropy}
Table \ref{tab:varaniso} and Fig.~\ref{fig:lambda_varaniso} show the
results for four sets of calculations in which the Coriolis number is
kept approximately constant at 0.3 whereas the turbulence anisotropy
$A_{\rm V}$ is varied between $-0.07$ and $-0.47$. This is done by
choosing suitable values for $f_0$ and $f_1$ so that the rms-velocity
stays approximately constant.

From the figure it is seen that the ratio of the stress to the amount
of anisotropy decreases monotonically as a function of $A_{\rm
  V}$. Although part of the difference can be explained by the
somewhat smaller Co (0.31 as opposed to 0.34 in the other cases) in
the calculation with the largest $A_{\rm V}$, the trend still
persists.

This trend can be understood as follows: from the approximate relation
$\Lambda_{\rm V} \approx 2\, \tauc A_{\rm V}$ we obtain
$\tauc \propto \Lambda_{\rm V}/A_{\rm V}$. The decreasing trend of
$\qyz/A_{V} \propto \Lambda_{\rm V}/A_{\rm V} \propto \tauc$ seen in
the results suggests that the correlation time changes when the
turbulence anisotropy is varied, i.e.\ $\tauc$ decreases when $A_{V}$
is increased. This is plausible since to change $A_{\rm V}$
different values of the forcing amplitudes, $f_0$ and $f_1$ (see
Eq.~\ref{equ:forceaniso}), need to be used resulting in differences in
the turbulence.

\begin{figure}[t]
\centering
\includegraphics[width=0.45\textwidth]{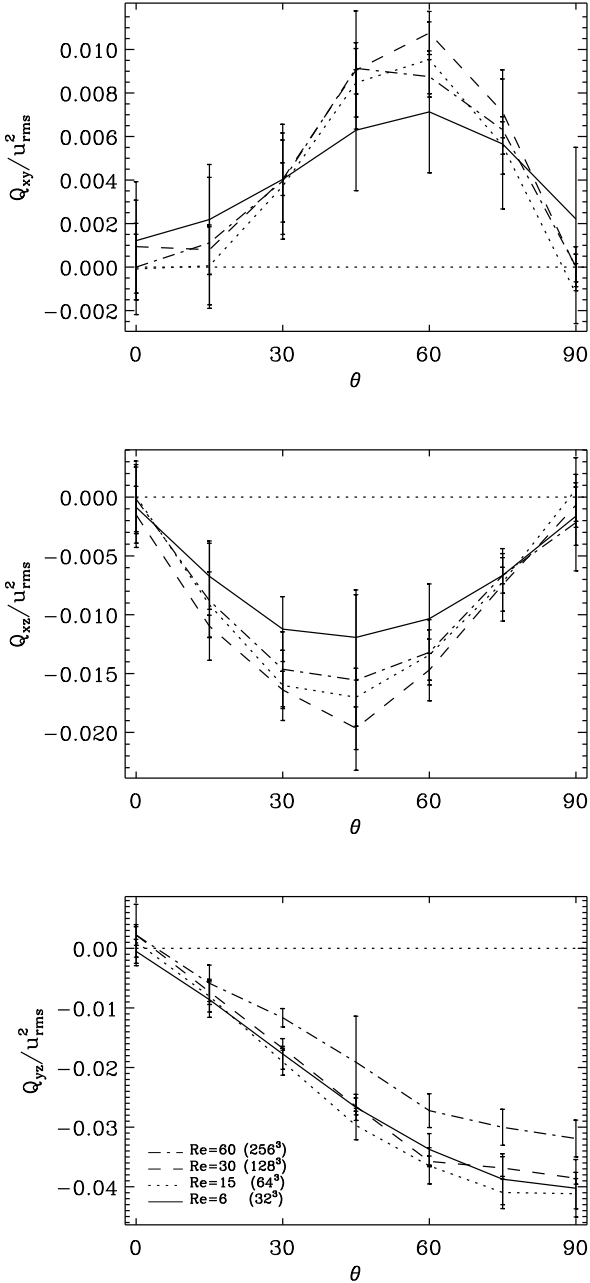}
\caption{Off-diagonal Reynolds stresses $Q_{xy}$ (top), $Q_{xz}$
  (middle), and $Q_{yz}$ (bottom) as functions of latitude and
  Reynolds number, see the legend in the lower-most panel.}
\label{fig:lambdaRe}
\end{figure}

\subsection{Dependence on Reynolds number}
Figure~\ref{fig:lambdaRe} shows the off-diagonal stresses as functions
of latitude and Reynolds number for a constant $\Co \approx 0.25$ (see also Table~\ref{tab:varRe}). 
There is no \emph{a priori} reason to
expect that the stresses should depend on the Reynolds number if
${\rm Re} \gg 1$ and if the turbulence anisotropy is kept constant. 
This is essentially what is borne out of the
simulations, although there seems to be a weak decreasing trend as a
function of Re for ${\rm Re} > 15$ for $\qxy$ and $\qxz$, although the
results are within error bars. For $\qyz$, however, the largest
Reynolds number case shows a distinct drop in comparison to the less
turbulent cases.

The decrease in the stresses as a function of the Reynolds number is
likely to have the same origin as the decrease seen when the
turbulence anisotropy is increased (see the previous section). In
order to obtain the same $A_{\rm V}$ in the simulations different
values, $f_0$ and $f_1$ are needed for different values of $\nu$, the
kinematic viscosity. More precisely, the less the viscosity, the
more difficult it is to obtain large anisotropy. Thus, for smaller
$\nu$, higher ratio $f_1/f_0$ is required to achieve a given
$A_{\rm V}$. Thus, if the interpretation that the higher the ratio
$f_1/f_0$, the smaller $\tauc$ becomes is correct, the decreasing
trend seen as a function of ${\rm Re}$ might be an artefact due to the
small differences in the forcing between the different runs.

\subsection{Strouhal number from simulations}
\label{subsec:validmta}
The MTA-model reproduces the simulation results for the off-diagonal
stresses reasonably well when $\St = 1 \ldots 2$ is used. Now we turn
to the numerical simulations to determine $\St$
independently.
Although we have not been able to provide any direct support of
the basic MTA-assumption, $T_{ij} = -\tau^{-1} Q_{ij}$,
the MTA-model is still able to reproduce many of the
features of the numerical simulations adequately. This implies that a
value for $\St$ could be extracted from the MTA-relations.

We consider two methods to determine the Strouhal number from the
simulations: (i) MTA-relations derived for the Reynolds stresses, and
(ii) similar relations for passive scalar transport under the
influence of rotation.

\onltab{3}{
   \begin{table*}
   \centering
   \caption[]{Turbulence anisotropies and Reynolds stresses for varying Reynolds numbers. $\kef/k_1=5$ and $\Omega_0 = 0.1$ were used in all runs. $\tilde{Q}_{ij} = Q_{ij}/\urms^2$.}
%      \vspace{-0.75cm}
      \label{tab:varRe}
     $$
         \begin{array}{p{0.055\linewidth}cccccccccrcrrr}
           \hline
           \noalign{\smallskip}
           Run      & {\rm Grid} & \theta & \Co & \nu [10^{-3}] & {\rm Re} & \urms & \tilde{Q}_{xx} & \tilde{Q}_{yy} & \tilde{Q}_{zz} & A_{\rm H} & A_{\rm V} & \tilde{Q}_{xy} & \tilde{Q}_{xz} & \tilde{Q}_{yz} \\
           \noalign{\smallskip}
           \hline
           \noalign{\smallskip}
           32c15  &  32^3 &  0\degr & 0.255 &   5 & 6.0 & 0.154 & 0.274 & 0.273 & 0.454 &-0.000 & -0.361 & 0.001 &-0.001 &-0.001 \\
           32c16  &  32^3 & 15\degr & 0.255 &   5 & 6.0 & 0.154 & 0.274 & 0.273 & 0.454 &-0.001 & -0.361 & 0.002 &-0.007 &-0.009 \\
           32c17  &  32^3 & 30\degr & 0.255 &   5 & 6.0 & 0.154 & 0.273 & 0.275 & 0.453 & 0.001 & -0.359 & 0.004 &-0.011 &-0.018 \\
           32c18  &  32^3 & 45\degr & 0.255 &   5 & 6.0 & 0.154 & 0.272 & 0.278 & 0.451 & 0.006 & -0.352 & 0.006 &-0.012 &-0.027 \\
           32c19  &  32^3 & 60\degr & 0.255 &   5 & 6.0 & 0.154 & 0.270 & 0.283 & 0.448 & 0.014 & -0.344 & 0.007 &-0.010 &-0.034 \\
           32c20  &  32^3 & 75\degr & 0.255 &   5 & 6.0 & 0.154 & 0.267 & 0.288 & 0.446 & 0.021 & -0.338 & 0.006 &-0.006 &-0.039 \\
           32c21  &  32^3 & 90\degr & 0.255 &   5 & 6.0 & 0.154 & 0.266 & 0.290 & 0.445 & 0.025 & -0.335 & 0.002 &-0.001 &-0.041 \\
           \hline
           \noalign{\smallskip}
           64e15  &  64^3 &  0\degr & 0.254 &   2 &  15 & 0.155 & 0.273 & 0.274 & 0.454 & 0.001 & -0.361 & 0.000 & 0.000 & 0.001 \\
           64e16  &  64^3 & 15\degr & 0.254 &   2 &  15 & 0.155 & 0.273 & 0.274 & 0.454 & 0.001 & -0.361 & 0.000 &-0.009 &-0.008 \\
           64e17  &  64^3 & 30\degr & 0.254 &   2 &  15 & 0.155 & 0.273 & 0.276 & 0.451 & 0.003 & -0.353 & 0.004 &-0.016 &-0.019 \\
           64e18  &  64^3 & 45\degr & 0.254 &   2 &  15 & 0.155 & 0.271 & 0.283 & 0.446 & 0.011 & -0.339 & 0.009 &-0.017 &-0.030 \\
           64e19  &  64^3 & 60\degr & 0.254 &   2 &  15 & 0.155 & 0.269 & 0.290 & 0.442 & 0.021 & -0.324 & 0.010 &-0.013 &-0.036 \\
           64e20  &  64^3 & 75\degr & 0.253 &   2 &  15 & 0.155 & 0.265 & 0.297 & 0.439 & 0.032 & -0.316 & 0.006 &-0.007 &-0.041 \\
           64e21  &  64^3 & 90\degr & 0.253 &   2 &  15 & 0.155 & 0.263 & 0.300 & 0.437 & 0.037 & -0.311 &-0.001 & 0.000 &-0.041 \\
           \hline
           \noalign{\smallskip}
           128b15 & 128^3 &  0\degr & 0.252 &   1 &  31 & 0.156 & 0.274 & 0.275 & 0.452 & 0.000 & -0.355 & 0.001 &-0.001 & 0.002 \\
           128b16 & 128^3 & 15\degr & 0.251 &   1 &  31 & 0.156 & 0.276 & 0.275 & 0.450 &-0.001 & -0.349 & 0.001 &-0.011 &-0.007 \\
           128b17 & 128^3 & 30\degr & 0.251 &   1 &  31 & 0.156 & 0.278 & 0.277 & 0.447 &-0.001 & -0.339 & 0.004 &-0.016 &-0.017\\
           128b18 & 128^3 & 45\degr & 0.251 &   1 &  31 & 0.156 & 0.278 & 0.283 & 0.440 & 0.006 & -0.320 & 0.009 &-0.019 &-0.026\\
           128b19 & 128^3 & 60\degr & 0.250 &   1 &  31 & 0.157 & 0.275 & 0.291 & 0.435 & 0.016 & -0.304 & 0.011 &-0.015 &-0.035 \\
           128b20 & 128^3 & 75\degr & 0.250 &   1 &  31 & 0.157 & 0.270 & 0.299 & 0.432 & 0.029 & -0.295 & 0.007 &-0.007 &-0.037 \\
           128b21 & 128^3 & 90\degr & 0.250 &   1 &  31 & 0.157 & 0.268 & 0.303 & 0.430 & 0.035 & -0.289 &-0.000 &-0.001 &-0.038 \\
           \hline
           \noalign{\smallskip}
           256a15 & 256^3 &  0\degr & 0.251 & 0.5 &  61 & 0.156 & 0.275 & 0.276 & 0.449 & 0.001 & -0.347 &-0.000 &-0.000 & 0.003 \\
           256a16 & 256^3 & 15\degr & 0.252 & 0.5 &  61 & 0.156 & 0.277 & 0.276 & 0.447 &-0.002 & -0.341 & 0.001 &-0.008 &-0.005 \\
           256a17 & 256^3 & 30\degr & 0.252 & 0.5 &  61 & 0.156 & 0.276 & 0.277 & 0.448 & 0.001 & -0.342 & 0.004 &-0.015 &-0.011  \\
           256a18 & 256^3 & 45\degr & 0.250 & 0.5 &  61 & 0.157 & 0.275 & 0.282 & 0.443 & 0.006 & -0.330 & 0.009 &-0.015 &-0.020 \\
           256a19 & 256^3 & 60\degr & 0.251 & 0.5 &  61 & 0.156 & 0.273 & 0.289 & 0.438 & 0.016 & -0.314 & 0.009 &-0.013 &-0.027 \\
           256a20 & 256^3 & 75\degr & 0.250 & 0.5 &  61 & 0.157 & 0.269 & 0.296 & 0.435 & 0.027 & -0.306 & 0.006 &-0.007 &-0.030 \\
           256a21 & 256^3 & 90\degr & 0.250 & 0.5 &  61 & 0.157 & 0.264 & 0.301 & 0.435 & 0.037 & -0.305 & 0.000 &-0.002 &-0.032 \\
           \hline
         \end{array}
     $$ 
   \end{table*}
}

\subsubsection{Off-diagonal Reynolds stresses versus MTA-relations}
\begin{figure}[t]
\centering
\includegraphics[width=0.5\textwidth]{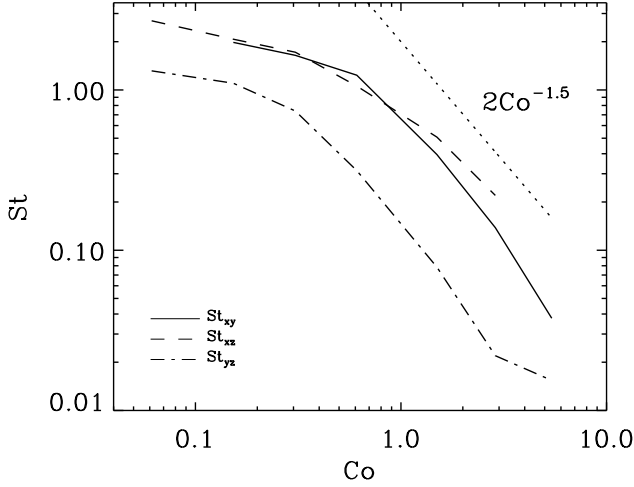}
\caption{Strouhal numbers computed from Eq.~(\ref{equ:Stij}) at
  colatitude $\theta=60\degr$ as functions of the Coriolis number for the runs listed in Table~\ref{tab:Re12}. A
  power law proportional to $\Co^{-1.5}$ shown for reference.}
\label{fig:pstrouhal}
\end{figure}

Using the minimal tau-approximation 
and assuming a stationary state where 
$\dot{\vec{U}} = \vec{f}_{\rm force} + \vec{f}_{\rm visc} = 0$, 
the off-diagonal
Reynolds stresses can be derived from Eq.~(\ref{equ:NS}), yielding
\begin{eqnarray}
\qxy & = & 2\,\Omz \tau_{xy} (\qyy-\qxx) + 2\,\Omx \tau_{xy} \qxz\;, \label{equ:MTAqxy} \\
\qxz & = & 2\,\Omz \tau_{xz} \qyz - 2\,\Omx \tau_{xz} \qxy\;, \label{equ:MTAqxz} \\
\qyz & = & 2\,\Omx \tau_{yz} (\qzz-\qyy) - 2\,\Omz \tau_{yz} \qxz\;, \label{equ:MTAqyz}
\end{eqnarray}
where $\tau_{ij}$ are a set of relaxation times that allow for
the possibility that there can be different values of $\tau$
for different components of the Reynolds stress.
Here $\qij$ are the components of the Reynolds
stress tensor in the simulations. These
equations can be solved for the $\tau_{ij}$ via
\begin{eqnarray}
\tau_{xy} & = & \frac{\qxy}{2\,\Omz(\qyy-\qxx) + 2\,\Omx\qxz} \;, \label{equ:mtatauxy} \\
\tau_{xz} & = & \frac{\qxz}{2\,\Omz \qyz - 2\,\Omx\qxy} \;, \label{equ:mtatauxz} \\
\tau_{yz} & = & \frac{\qxy}{2\,\Omx(\qzz-\qyy) + 2\,\Omz\qxz} \;. \label{equ:mtatauyz}
\end{eqnarray}
The first and the last of these equations, i.e. $\tau_{xy}$ and
$\tau_{yz}$, yield reasonable values for most cases,
whereas $\tau_{xz}$ is less well-behaved,
producing sign changes and showing no clear trend in magnitude 
as a function of latitude or rotation. This is because 
the two terms in the denominator of Eq.~(\ref{equ:mtatauxz}) tend to nearly cancel 
each other; see the upper panels of Fig.~\ref{fig:lambda}.
If, however, the values of $\tau$ in
Eqs.~(\ref{equ:MTAqxy}) to (\ref{equ:MTAqyz}) are considered the same,
which is a basic MTA-assumption,
it is possible to solve for $\qxz$ in terms of the diagonal stresses
\begin{eqnarray}
\qxz = \frac{4\,\Omx \Omz \tau_{xz}^2}{1+4\,\Omega_0^2 \tau_{xz}^2}(\qxx + \qzz - 2\,\qyy)\;, \label{equ:mtaqxz}
\end{eqnarray}
from which it follows that
\begin{eqnarray}
\tau_{xz} = \bigg[\frac{\qxz}{4\,\Omx \Omz (\qxx + \qzz -2\,\qyy) - 4\,\Omega_0^2 \qxz} \bigg]^{1/2}. \label{equ:mtatauxz2}
\end{eqnarray}
Although the assumption that all $\tau_{ij}$ are equal in
Eqs.~(\ref{equ:MTAqxy}) to (\ref{equ:MTAqyz}) is 
not exactly realized in the numerical simulations, 
relation
(\ref{equ:mtatauxz2}) gives results that are compatible with those
from Eqs.~(\ref{equ:mtatauxy}) and (\ref{equ:mtatauyz}). The Strouhal
number can now be computed from
\begin{eqnarray}
\St_{ij} = \tau_{ij} \urms \kef\;, \label{equ:Stij}
\end{eqnarray}
where $\tau_{ij}$ are given by Eqs.~(\ref{equ:mtatauxy}),
(\ref{equ:mtatauxz2}), and (\ref{equ:mtatauyz}),
respectively. Representative results are given in
Fig.~\ref{fig:pstrouhal}.

\subsubsection{Passive scalar transport with rotation}
As an independent check of the dependence of the Strouhal number on
rotation, we expand the passive scalar transport case, which was
studied by Brandenburg et al.\ (\cite{Brandenburgea2004})
to cases where rotation is included. The numerical model is the same
as in the runs presented so far, except that isotropic forcing is used
with $f_0=0.01-0.03$ and $f_1 = 0$ in
Eq.~(\ref{equ:forceaniso}). 

The turbulent passive scalar flux is denoted by
$\overline{\mathcal{F}}_i = \overline{u_i c}$, where $c$ is the
fluctuation of passive scalar density, i.e.\ the passive scalar
concentration per unit volume. Following the MTA-approach, we
solve first for the time derivative
\begin{equation}
\dot{\overline{\mathcal{F}}}_i = \overline{\dot{u}_i c} + \overline{u_i \dot{c}}\;,
\end{equation}
where the fluctuation of the passive scalar field, neglecting
diffusive terms, is given by
\begin{equation}
\dot{c} = - \vec{\nabla} \cdot (\vec{u} \overline{C} + \overline{\vec{U}} c + \vec{u} c) \;.
\end{equation}
Here $\overline{C}$ and $\overline{\vec{U}}$ are the mean passive
scalar concentration and the mean velocity, respectively. Following
Brandenburg et al.\ (\cite{Brandenburgea2004}) we impose a large-scale
gradient of passive scalar concentration according to $\vec{\nabla}
\overline{C} = (0,0,G)^T$. In what follows $G=0.1$ is used. 
Now, assuming incompressibility and using
the fact that $\overline{\vec{U}} = 0$, we arrive at
\begin{equation}
\dot{\overline{\mathcal{F}}}_i = - \qij \partial_j \overline{C} - 2\,\varepsilon_{ijk} \Omega_j \overline{\mathcal{F}}_k - T_i^{\rm (1)} - T_i^{\rm (2)} - T_i^{\rm (3)}\;,\label{equ:dotFi}
\end{equation}
where the last three terms denote the triple correlations. For the
$z$-component of Eq.~(\ref{equ:dotFi}) the triple correlations are
given by
\begin{eqnarray}
T_z^{\rm (1)} = \overline{u_z {\bm \nabla}\cdot ({\bm u} c)}\;, \;\; T_z^{\rm (2)} = \overline{ ({\bm u} c) \cdot \bm{\nabla} u_z}\;, \;\; T_z^{\rm (3)} = \overline{c\nabla_z h}, 
\end{eqnarray}
where $h = \cst \ln \rho$ is the reduced pressure (or enthalpy). In the non-rotating
case, $T_z^{\rm (1)} + T_z^{\rm (2)} =0$, and the contributions from
the momentum equation also cancel on average, i.e.\ $T_z^{\rm (2)} +
T_z^{\rm (3)} =0$. The former relation follows from the periodic
boundary conditions used and remains valid when rotation is added. The
latter relation, however, is no longer true and $T_z^{\rm (3)}$ is now
balanced by the Coriolis term. Thus, the MTA-assumption should be
applied to $T_z^{\rm (3)}$
\begin{equation}
\overline{c \nabla_z h} = \tau^{-1} \overline{\mathcal{F}}_z\;.
\end{equation}
Assuming a stationary state in Eq.~(\ref{equ:dotFi}), the passive scalar
fluxes can now be written as
\begin{eqnarray}
\overline{\mathcal{F}}_x &=& -\tau_x \qxz G + 2\,\Omz \tau_x \overline{\mathcal{F}}_y\;, \label{equ:pscalarfx} \\
\overline{\mathcal{F}}_y &=& -\tau_y \qyz G - 2\,\Omz \tau_y \overline{\mathcal{F}}_x + 2\,\Omx \tau_y \overline{\mathcal{F}}_z\;, \label{equ:pscalarfy} \\
\overline{\mathcal{F}}_z &=& -\tau_z \qzz G - 2\,\Omx \tau_z \overline{\mathcal{F}}_y\;, \label{equ:pscalarfz}
\end{eqnarray}
where we have retained the possibility that the values of $\tau$ from different
equations are unequal.

In the passive scalar cases we consider isotropically forced
turbulence for which $\qxz \approx \qyz \approx 0$ even when rotation
is included. Equations (\ref{equ:pscalarfx})--(\ref{equ:pscalarfz}) yield
\begin{eqnarray}
\tau_x &=& \frac{\overline{\mathcal{F}}_x}{2\,\Omz \overline{\mathcal{F}}_y}\;, \label{equ:pscalartaux} \\
\tau_y &=& \frac{\overline{\mathcal{F}}_y}{2\,\Omx \overline{\mathcal{F}}_z - 2\,\Omz \overline{\mathcal{F}}_x}\;, \label{equ:pscalartauy} \\
\tau_z &=& \frac{-\overline{\mathcal{F}}_z}{\qzz G + 2\,\Omx \overline{\mathcal{F}}_y}\;,\label{equ:pscalartauz}
\end{eqnarray}
which can be used to compute the Strouhal numbers
\begin{eqnarray}
\St_i = \tau_i \urms \kef, \label{equ:Stij_pscalar}
\end{eqnarray}
where $\tau_i$ are given by Eqs.~(\ref{equ:pscalartaux}) to
(\ref{equ:pscalartauz}).

\begin{figure}[h!]
\centering
\includegraphics[width=0.45\textwidth]{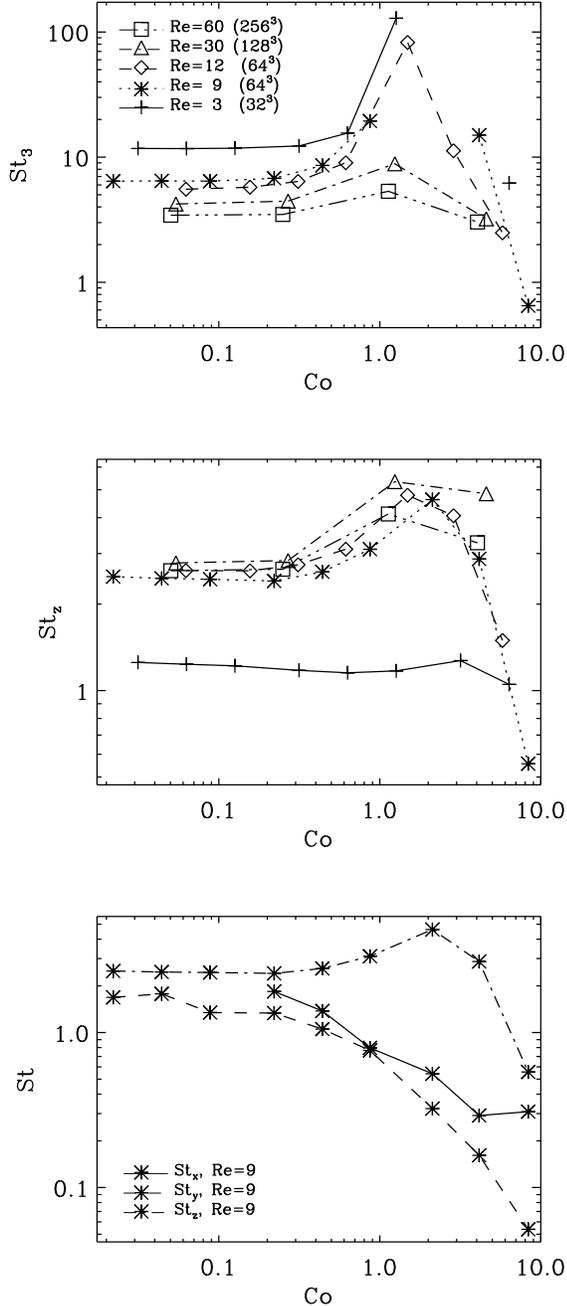}
\caption{Uppermost panel: St from triple correlation $T_z^{\rm
    (3)}$, middle panel: $\St_z$, corresponding to the tau in
  Eq.~(\ref{equ:pscalartauz}), lowermost panel: Strouhal numbers from
  Eq.~(\ref{equ:Stij_pscalar}) as functions of the Coriolis number for
  a constant Reynolds number of roughly nine.}
\label{fig:pstrouhal_pscalar}
\end{figure}

In the passive scalar case, the second and third order terms are
indeed correlated (Brandenburg et al.\ \cite{Brandenburgea2004}) 
according to the basic MTA-assumption, and a
Strouhal number can be thus computed using
\begin{equation}
\St_3 = \tau_3 \urms \kef\;,\label{equ:St3}
\end{equation}
where 
\begin{equation}
\tau_3 = \overline{\mathcal{F}}_z\left/\,\overline{c \nabla_z h}\right.\;.
\end{equation}
See Fig.~\ref{fig:pstrouhal_pscalar} for representative results.

\subsubsection{Discussion}
For slow rotation, the Strouhal number is consistently 
between one and three when
computed from the Reynolds stresses (Fig.~\ref{fig:pstrouhal}). 
Similar values are obtained from the passive scalar transport
(Fig.~\ref{fig:pstrouhal_pscalar}) when the Reynolds number is
ten or larger. The Strouhal number computed from the triple
correlations, $\St_3$, is more strongly dependent on the Reynolds number, 
but it
seems to converge slowly towards a constant value near
unity. These results are in line with the values required in the
MTA-model and earlier studies in different contexts
employing similar turbulence calculations (Brandenburg et
al.\ \cite{Brandenburgea2004}; Brandenburg \& Subramanian
\cite{BranSubra2005}, \cite{BranSubra2007}).

However, when the Coriolis number approaches or exceeds unity, the
Strouhal numbers computed from the equations of the Reynolds stresses,
Eqs.~(\ref{equ:mtatauxy})--(\ref{equ:mtatauyz}), decrease rapidly so
that for $\Co \approx 5$ it has dropped at least by an order
of magnitude (see Fig.~\ref{fig:pstrouhal}). 
Similar results are obtained for the passive scalar transport under
the influence of rotation, see
Eqs.~(\ref{equ:pscalartaux})--(\ref{equ:pscalartauz}) and the lower
panels of Fig.~\ref{fig:pstrouhal_pscalar}. The trend is clearer
for $\St_x$ and $\St_y$, whereas for $\St_z$ the decreasing trend is
seen only for rapid rotation, i.e.\ when $\Co > 2\ldots3$.  For slow
rotation, however, $\St_z$ is almost constant and increases somewhat
when the Coriolis number approaches unity. These results seem to
confirm the trend seen earlier in convection simulations (K\"apyl\"a
et al.\ \cite{Kaepylaeea2005}, \cite{Kaepylaeea2006a}).

The Strouhal number from the triple correlations follows a trend
similar to $\St_z$, with increasing values up to $\Co \approx
2\ldots3$ after which there is a rapid decrease, see the uppermost
panel of Fig.~\ref{fig:pstrouhal_pscalar}. For low Reynolds number St
can become negative in the range $\Co = 1 \ldots 5$, hence the gaps in
the corresponding data in Fig.~\ref{fig:pstrouhal_pscalar}.

%______________________________________________________________

\section{Conclusions}
\label{sec:conclusions}
Turbulent momentum fluxes, which are described by the Reynolds stresses, were
determined from numerical simulations of homogeneous rotating
anisotropic turbulence. Since no large-scale shear is present, the
generated Reynolds stresses correspond to contributions that are 
already present for uniform rotation.
The resulting term is known as the $\Lambda$-effect (Krause \& R\"udiger
\cite{KrauseRued1974}). The component responsible for the horizontal
transport, $\qxy$, is positive and peaks around latitude $30\degr$
regardless of the Coriolis number. The vertical component is
predominantly negative and it always peaks at the equator. 

Although the numerical results for the $\Lambda$-effect broadly
agree with analytical SOCA calculations (Kitchatinov \& R\"udiger
\cite{KitRued1993}, \cite{KitRued2005}), the MTA-model seems to
reproduce certain features of the numerical results somewhat more closely.
The present
numerical results do not show the enigmatic results, such as the extreme
latitude distribution of $\qxy$ or a positive $\qyz$ for rapid rotation,
which have been reported from convection simulations (e.g.\ Chan
\cite{Chan2001}; K\"apyl\"a et al.\ \cite{Kaepylaeea2004}). The
difference lies most likely in our 
neglecting stratification and heat fluxes. The exact manner in
which they affect the Reynolds stresses is not within the scope of the
present paper, but should be investigated more closely in the future.

By applying the minimal tau approximation closure relation to the
Reynolds stress equation, qualitatively similar results are obtained,
but the magnitude of the stresses is in general too large. The
vertical flux in the MTA-model, however, has a maximum at
mid-latitudes for intermediate and rapid rotation. Adding an empirical
rotational isotropization term (motivated in Sect.~\ref{sec:diaresults})
also brings the magnitude in line with
the 3D simulations. Although adding this term with this particular
form has no rigorous theoretical basis, we can see that phenomenological
effects of isotropization of turbulence due to rotation are indeed seen in the
simulations and that the term is thus justified.

Another drawback of the MTA-model is that the diagonal components of
the Reynolds tensor are rather badly reproduced since the nonlinear
effects of rotation manifest in the numerical simulations are not
explicitly taken into account. The empirically added rotational
isotropization term augments the magnitudes, but not the latitude
distribution. Furthermore, no direct evidence of the validity of the
MTA-assumption $\qij = -T_{ij}/\tau$ was found in the numerical
simulations. Contrasting the behavior of the diagonal components to
the fairly good correspondence between the numerical simulations and
the MTA-model for the off-diagonal components leads us to conclude
that, where the behavior of the diagonal components is dominated by
the inadequately modeled nonlinear effects, the off-diagonals are
fairly well presented by the linear terms.

A Strouhal number of order unity in the MTA-model gives best fits
to the numerical results. Fitting the numerical results to expressions
derived under the MTA, similar values of $\St$ are found for slow rotation. For
Coriolis numbers of order unity or larger, however, the
Strouhal number obtained in this manner decreases rapidly.
In the passive scalar case, the situation is somewhat more complex,
although a similar decreasing trend of the Strouhal number is recovered for rapid
rotation, see Fig.~\ref{fig:pstrouhal_pscalar}.
These results are in accordance
with earlier results from convection simulations (K\"apyl\"a et al.\
\cite{Kaepylaeea2005}, \cite{Kaepylaeea2006a}) using Reynolds
stresses or correlation analysis of the velocity field. 

A related aspect in turbulent transport that requires closer
study is the turbulent viscosity (see preliminary results in
K\"apyl\"a \& Brandenburg \cite{KaBr2007}) and the possibility of a
$\Lambda$-effect due to the anisotropy induced by a large-scale shear
flow (Leprovost \& Kim \cite{LeproKim2007}). These matters will be
considered in more detail in a future publication.

\begin{acknowledgements}
  The computations were performed on the facilities hosted by the
  Center of Scientific Computing in Espoo, Finland, who are financed
  by the Finnish ministry of education. PJK acknowledges the financial
  support from Helsingin Sanomat foundation and the Academy of Finland
  grant No.\ 121431. PJK acknowledges the hospitality of Nordita
  during the program `Turbulence and Dynamos' during which this work
  was finalized. The anonymous referee is acknowledged for the
  critical reading and helpful comments on the manuscript.
\end{acknowledgements}

\appendix

%______________________________________________________________

\section{Stationary solutions for the stresses from MTA}
\label{app:mtaeqs}
Setting $\dot{Q}_{ij} = 0$ and using the MTA-closure $T_{ij} =
-Q_{ij}/\tau$, and parameterizing the contributions of the forcing as
$\overline{u_i f_j} + \overline{u_j f_i} = Q_{ij}^{(0)}/\tau$ in
Eq.~(\ref{equ:mtastress}) gives six equations for the six unknowns of
the symmetric tensor $Q_{ij}$, i.e.\
\begin{eqnarray}
\qxx & = & 4\,\Omz \tau \qxy + \qxx^{(0)}\;, \label{equ:appqxx}\\  
\qxy & = & 2\,\Omz \tau (\qyy-\qxx) + 2\,\Omx \tau \qxz\;, \\  
\qxz & = & 2\,\Omz \tau \qyz - 2\,\Omx \tau \qxy\;, \\  
\qyy & = & 4\,\Omx \tau \qyz - 4\,\Omz \tau \qxy + \qyy^{(0)}\;, \label{equ:appqyy}\\  
\qyz & = & 2\,\Omx \tau (\qzz-\qyy) - 2\,\Omz \tau \qxz\;, \\  
\qzz & = & -4\,\Omx \tau \qyz + \qzz^{(0)}\;. \label{equ:appqzz}
\end{eqnarray}
It is possible to solve for $Q_{ij}$ in terms of $\tau$, $\Omx$,
$\Omz$, and $Q_{ij}^{(0)}$. From Eqs.~(\ref{equ:appqxx}) to
(\ref{equ:appqzz}) it is clear that only three components of $Q_{ij}$
are independent. Thus it is sufficient to solve for the off-diagonal
components. After some algebra we arrive at
\begin{eqnarray}
\qxy &=& 2\, \Omz \tau \bigg( \frac{K_0}{K_1} \bigg) (\qyy^{(0)} - \qxx^{(0)}) + \nonumber \\ && \hspace{2cm} 24\,\Omx^2 \Omz \tau^3 K_0 (\qzz^{(0)} - \qyy^{(0)})\;, \label{equ:mtamodelqxy}\\
\qxz &=& 2\,\Omz \tau \qyz - 2\,\Omx \tau \qxy\;, \\  
\qyz &=& 2\,\Omx \tau K_1 (1 + 144\,\Omx^2\Omz^2 \tau^4 K_0) (\qzz^{(0)} - \qyy^{(0)}) + \nonumber \\ && \hspace{2cm} 24\,\Omx \Omz^2 \tau^3 K_0 (\qyy^{(0)} - \qxx^{(0)}) \;,
\end{eqnarray}
where 
\begin{eqnarray}
K_0 &=& \frac{1}{1+20\,\Omega_0^2 \tau^2 + 64\,\Omega_0^4 \tau^4} \equiv \nonumber \\ && \hspace{3cm} \frac{1}{1+5\Co_{\rm MTA}^2+8\Co_{\rm MTA}^4}, \\
K_1 &=& \frac{1}{1+4\,\Omz^2\tau^2+16\,\Omx^2\tau^2} \equiv \nonumber \\ && \hspace{3cm} \frac{1}{1+\Co_{\rm MTA}^2(1+3\sin^2\theta)}\;,
\end{eqnarray}
and $\Co_{\rm MTA}$ is given by Eq.~(\ref{equ:CoMTA}).

In the present study the forcing is such that $\qyy^{(0)} - \qxx^{(0)}
= 0$, so the equations reduce to
\begin{eqnarray}
\qxy &=& 24\,\Omx^2 \Omz \tau^3 K_0 (\qzz^{(0)} - \qyy^{(0)})\;,\\
\qxz &=& 2\,\Omz \tau \qyz - 2\,\Omx \tau \qxy\;, \\  
\qyz &=& 2\,\Omx \tau K_1 (1 + 144\,\Omx^2\Omz^2 \tau^4 K_0) (\qzz^{(0)} - \qyy^{(0)}).
\end{eqnarray}

\section{Coefficients of the $\Lambda$-effect from SOCA}
\label{sec:socalambda}
The fluxes of angular momentum have commonly been parameterized by
Eqs.~({\ref{equ:conqxy}) and ({\ref{equ:conqyz}), and the normalized
fluxes by Eqs.~(\ref{equ:H}) and (\ref{equ:V}). Kitchatinov \&
R\"udiger (\cite{KitRued2005}) computed these coefficients using
SOCA
\begin{eqnarray}
H &=& H^{(1)}(\Cost) \sin^2 \theta\;, \\
V &=& V^{(0)}(\Cost) - H^{(1)}(\Cost) \cos^2 \theta\;,
\end{eqnarray}
where $\Cost = 2\,\Omega_0 \tauto$ is their definition of the
Coriolis number, and $\tauto$ the turnover time. Note that there is
a difference of $2\pi$ in comparison to our definition,
Eq.~(\ref{equ:Coriolis}). For simplicity, we retain $\Cost$
in the expressions that follow. The coefficients $H^{(1)}$ and $V^{(0)}$ are given by
\begin{eqnarray}
H^{(1)} &=& \bigg(\frac{l_{\rm corr}}{H_\rho}\bigg)^{2} [J_1(\Cost) + aI_1(\Cost)]\;, \\ 
V^{(0)} &=& \bigg(\frac{l_{\rm corr}}{H_\rho}\bigg)^{2} [J_0(\Cost) + aI_0(\Cost)]\;,
\end{eqnarray}
where $l_{\rm corr}$ is the correlation length, $H_\rho$ the
density scale height, and $a = 2$ an `anisotropy parameter' that
reduces the amount of anisotropy for slow rotation.

The functions $I_i$ and $J_i$ are given by
\begin{eqnarray}
I_0 &=& \frac{1}{4\,\Cost^2} \bigg(-19 - \frac{5}{1+\Cost^2} + \frac{3\,\Cost^2 + 24}{\Cost} \arctan \Cost \bigg), \\
I_1 &=& \frac{3}{4\,\Cost^2} \bigg(-15 - \frac{2\,\Cost^2}{1+\Cost^2} + \frac{3\,\Cost^2 + 15}{\Cost} \arctan \Cost \bigg), \\
J_0 &=& \frac{1}{2\,\Cost^4} \bigg(9 - \frac{2\,\Cost^2}{1+\Cost^2} - \frac{\Cost^2 + 9}{\Cost} \arctan \Cost \bigg), \\
J_1 &=& \frac{1}{2\,\Cost^4} \bigg(45 + \Cost^2 - \frac{4\,\Cost^2}{1+\Cost^2} + \nonumber \\  &&\hspace{2.5cm} \frac{\Cost^4 - 12\,\Cost^2 - 45}{\Cost} \arctan \Cost \bigg).
\end{eqnarray}
In our model there is no density stratification, so
$l_{\rm corr}/H_\rho$ is taken to be a free parameter that we choose
to be equal to unity.

\end{document}